\documentclass[12pt]{article}

\usepackage{natbib}
\usepackage{setspace}
\usepackage[left=1in,top=1in,right=1in,bottom=1in]{geometry}
\usepackage{graphicx}
\usepackage{amssymb}
\usepackage{amsmath}
\usepackage{theorem}
\usepackage{version}
\usepackage[dvipsnames]{xcolor}
\usepackage[normalem]{ulem}
\usepackage{multirow}
\usepackage{multicol}
\usepackage{epstopdf}
\usepackage{xcolor}
\usepackage{subcaption}
\usepackage{floatrow}
\usepackage{enumitem}
\usepackage{tikz}
\usetikzlibrary{arrows,arrows.meta,decorations,decorations.pathreplacing,calc,matrix}

\definecolor{Red}{rgb}{1,0,0}
\definecolor{Blue}{rgb}{0,0,1}
\definecolor{Green}{rgb}{0,1,0}
\definecolor{magenta}{rgb}{1,0,.6}
\definecolor{lightblue}{rgb}{0,.5,1}
\definecolor{lightpurple}{rgb}{.6,.4,1}
\definecolor{gold}{rgb}{.6,.5,0}
\definecolor{orange}{rgb}{1,0.4,0}
\definecolor{hotpink}{rgb}{1,0,0.5}
\definecolor{newcolor2}{rgb}{.5,.3,.5}
\definecolor{newcolor}{rgb}{0,.3,1}
\definecolor{newcolor3}{rgb}{1,0,.35}
\definecolor{darkgreen1}{rgb}{0, .35, 0}
\definecolor{darkgreen}{rgb}{0, .6, 0}
\definecolor{darkred}{rgb}{.75,0,0}
\definecolor{lightgrey}{rgb}{.7,.7,.7}

\definecolor{clemson-orange}{RGB}{234,106,32}
\definecolor{chicago-maroon}{RGB}{128,0,0}
\definecolor{northwestern-purple}{RGB}{82,0,99}
\definecolor{cornell-red}{RGB}{179,27,27}
\definecolor{sauder-green}{RGB}{171,180,0}
\definecolor{lawngreen}{RGB}{0,250,154}

\usepackage[unicode=true,
 bookmarks=true,bookmarksnumbered=true,bookmarksopen=true,bookmarksopenlevel=4,
 breaklinks=false,pdfborder={0 0 1},backref=false,colorlinks=true]
 {hyperref}
\hypersetup{ pdfborderstyle={},urlcolor=chicago-maroon,linkcolor=chicago-maroon,citecolor=northwestern-purple}

\usepackage{times}
\newtheorem{assumption}{Assumption}

\newtheorem{example}{Example}

\newtheorem{lemma}{Lemma}

\newtheorem{proposition}{Proposition}

{\theorembodyfont{\normalfont}

}

\newenvironment{proof}[1][Proof]{\textbf{#1.} }{\hfill \rule{0.5em}{0.5em} \bigskip}

\usepackage[nameinlink]{cleveref}
\crefname{assumption}{Assumption}{Assumptions}
\crefname{lemma}{Lemma}{Lemmas}
\crefname{theorem}{Theorem}{Theorems}
\crefname{corollary}{Corollary}{Corollaries}
\crefname{proposition}{Proposition}{Propositions}
\crefname{claim}{Claim}{Claims}
\crefname{procedure}{Procedure}{Procedures}
\crefname{algorithm}{Algorithm}{Algorithms}
\crefname{figure}{Figure}{Figures}
\crefname{remark}{Remark}{Remarks}
\crefname{section}{Section}{Sections}
\crefname{procedure}{Procedure}{Procedures}
\crefname{example}{Example}{Examples}
\crefname{definition}{Definition}{Definitions}
\crefname{table}{Table}{Tables}
\crefname{equation}{}{}
\crefname{enumi}{}{}
\crefname{conjecture}{Conjecture}{Conjectures}
\crefname{step}{Step}{Steps}
\crefname{appendix}{Appendix}{Appendices}
\crefname{footnote}{Footnote}{Footnotes}

\begin{document}

\title{Persuasion in Veto Bargaining\footnote{We are grateful to Nageeb Ali, Navin Kartik, John Patty, Maggie Penn, Carlo Prato, Ian Turner and seminar audiences at Emory, University of Pittsburgh, and UC Riverside for helpful comments. Kyungmin Kim is supported by the Ministry of Education of the Republic of Korea and the National Research Foundation of Korea (NRF-2020S1A5A2A03043516).}}
\author{Jenny S. Kim\thanks{Department of Political Science, Emory University, Email: seoyeon.kim@emory.edu}\hspace{12mm}Kyungmin Kim\thanks{Department of Economics, Emory University, Email: kyungmin.kim@emory.edu}\hspace{12mm}Richard Van Weelden\thanks{Department of Economics, University of Pittsburgh, rmv22@pitt.edu.}}
\date{\today}
\maketitle

\begin{abstract}
We consider the classic veto bargaining model but allow the agenda setter to engage in persuasion to convince the veto player to approve her proposal. 
We fully characterize the optimal proposal and experiment when Vetoer has quadratic loss, and show that the proposer-optimal can be achieved either by providing no information or with a simple binary experiment. Proposer chooses to reveal partial information when there is sufficient expected misalignment with Vetoer. In this  case the opportunity to engage in persuasion strictly benefits Proposer and increases the scope to exercise agenda power.  
\end{abstract}


\thispagestyle{empty}



\let \markeverypar \everypar
\newtoks \everypar
\everypar \markeverypar
\markeverypar{\the \everypar \looseness=-2\relax}
\newpage
\setcounter{page}{1}

\section{Introduction}
\label{s:intro}

There are many settings in which an agenda setter makes a proposal and other player(s) can decide whether to approve or reject (veto) it.  In almost all such cases, before the approval decision, the proposer attempts to persuade the veto player to approve the proposal they put forward.  For example, if a committee makes a recommendation (such as to hire a particular candidate) it will also present the case to get its recommendation approved.  Similarly, a legislator who introduces a bill will make a speech urging their colleagues to vote for it.  And, if the bill is passed by Congress, the Speaker or other Congressional leaders will make the case to the public, and the President, that the bill should not be vetoed. In this paper we explore how the possibility of persuasion augments agenda-setting power and influences the proposals that will be made.

Following the seminal contribution of \citet{RR:78}, a large literature has explored the consequences of being able to set the agenda.\footnote{See \citet{CM:04} for an excellent survey of the literature.} We extend the model to allow Proposer to also engage in Bayesian persuasion \citep{KG2011} to convince Vetoer to accept the proposal. We assume that Vetoer is uncertain of his optimal policy, and Proposer can design a Blackwell experiment to reveal information about it to Vetoer.  We consider both the case in which Proposer first engages in persuasion and then makes a proposal after observing the results, and the case in which Proposer makes a proposal without observing the outcome of persuasion.

As a leading example, consider an interest group who both develops legislation and lobbies for its passage.   Beyond campaign contributions and financial favors, interest groups can influence legislation both by providing a legislative subsidy \citep{HD:06}, and through informative lobbying \citep{AS:93, PvW:92, Schnakenberg:17, 
schnakenberg2021helping, schnakenberg2023formal, awad2020persuasive, ellis2020strategic, HKMY:23, dellis2023legislative}.  We interpret the former as the interest group using its resources and expertise to draft legislation and the latter as the interest group using information design to persuade the decisive policymaker---which could be the median legislator, the relevant committee chair, the executive, or even a critical bureaucrat.\footnote{\cite{PT:23} also consider persuasion of a bureaucrat.  In their model the bureaucrat can choose from an exogenous set of alternatives; in our framework the bureaucrat would choose between implementing new legislation or dragging their feet in which case the status quo is maintained.}  Policymakers act on a large number of issues, and outside of the central, highly salient ones, an individual policymaker may not know much about the issue at hand or have a developed position.  The interest group however can engage in informational lobbying which reveals (partial) information about the costs and benefits to the policymaker or their constituents, allowing them to learn about their most preferred policy.\footnote{Our informational environment is similar to that of \cite{HKMY:23}, in which a policy generates a known benefit to an interest group but an uncertain benefit to the policymaker. Relevant applications they discuss include tariff policy, travel restrictions to a particular country, and domestic environmental regulation.  Our paper differs in that the interest group can choose both the content of the proposal and the persuasion, whereas in their case the policy is fixed and the policymaker can learn about it at a cost.}  Based on this they decide whether or not to pass the legislation.\footnote{Policy development is costly, and so policymakers cannot simply adjust the policy content of the legislation \citep{HS:15}, but rather can decide whether to approve or reject it.}  In this setting, the interest group is the proposer and the policymaker the veto player.  

We consider bargaining between a proposer and a single veto player over a one dimensional policy. There is a status quo policy $0$ and Proposer's ideal policy is $1$. Vetoer's ideal point is $\theta \in [\underline{\theta}, \bar{\theta}]$, which is initially unknown to both players, but can be learned about from an experiment designed by Proposer. This can be interpreted as Vetoer's ideal action depending on an unknown state of the world, about which Proposer can reveal information.  In the bulk of our analysis we focus on the case in which Vetoer's loss is quadratic in the distance between his bliss point, $\theta$, and the implemented policy.  In this case, Vetoer is willing to accept a proposal of $p$ if and only if it is closer to his expected bliss point than the status quo of $0$. Exploiting this property and applying the state-of-the-art verification technique by \citet{DM:19}, we fully characterize Proposer's optimal proposal and experiment. 

We show that Proposer can achieve her maximal payoff by either providing no information or designing a simple binary experiment.  When Vetoer's bliss point is always positive ($\underline{\theta} \geq 0$), and so Proposer and Vetoer are always on the same side of the status quo, Proposer's payoff is maximized by providing no information.  This is because the resulting policy is proportional to Vetoer's expected bliss point, so when Proposer's utility is concave over policy she never benefits from revealing information. When Vetoer's bliss point can be negative $(\underline{\theta}<0)$, however, Proposer may benefit from revealing whether $\theta$ is above or below some threshold.  The key is that Proposer's agenda-setting authority ensures the resulting policy to be at least the status quo regardless of Vetoer preferences.  Consequently, by revealing when Vetoer's bliss point is on the opposite side of the status quo, Proposer can increase the expected policy.

We study two versions of the problem: the first in which Proposer runs an experiment prior to making a proposal (persuasion first) and the other where Proposer cannot condition on the experiment outcome (proposal first).\footnote{Whether persuasion and the proposal happen simultaneously or sequentially is not important; what matters is whether Proposer is able to observe what Vetoer learns prior to making her proposal.}  Whether persuasion precedes the proposal or not depends on the application.  In some cases, Proposer may be able to commission and reveal the outcome of a study or an opinion poll prior to engaging in policy bargaining.  In others---such as when proposing a particular candidate to be hired or confirmed---the proposal would necessarily come first.  

Under persuasion first, Proposer observes the realization of the experiment and can tailor her proposal to this realization, in contrast to proposal-first in which the realization is Vetoer's private information. Therefore, Proposer always does weakly better under persuasion first than proposal first.  However we show that Proposer does not strictly benefit from persuasion first; rather, the Proposer's solution is the same in both settings. As Proposer and Vetoer are symmetrically informed under persuasion first, Proposer proposes twice the conditional expectation of Vetoer's bliss point (or the status quo if negative) which is then accepted.  Analogous to the optimality of no information when Vetoer's bliss point is always positive, Proposer cannot benefit from inducing more than one strictly positive proposal: she could always do at least as well by pooling the signals and making a proposal based on the mean.  Hence, Proposer optimally generates a binary experiment, with the status quo maintained after one signal realization.  Notice that an identical outcome can be achieved under proposal first by making the same take-it-or-leave-it proposal followed by the same binary experiment.  Thus, the ability to tailor the proposal to the realization of the experiment provides no benefit to Proposer.

The optimal binary experiment is informative whenever the lower bound on Vetoer's bliss point is sufficiently negative relative to its expected value.  Intuitively, when Vetoer's expected bliss point is high, Proposer can get close to her ideal point without providing information and so chooses not to gamble by revealing information.  How much proposer is willing to reveal depends on her risk preferences and the distribution.  As Proposer's utility function becomes more concave she compromises more, resulting in a lower policy proposal and fewer vetoes. A likelihood ratio increase in the distribution of Vetoer bliss points also results in fewer vetoes but with a higher policy proposal.

When Vetoer does not have quadratic loss, proposal-first and persuasion-first are not always equivalent, and Proposer may do strictly better under persuasion-first.   Further, because more than just the expectation matters, it may be optimal to provide information even when Vetoer's bliss point is known to be positive.  We illustrate these points by considering a version in which Vetoer has a linear loss function, focusing on the case in which his ideal point could take one of two positive values.    

Our paper belongs to a large literature on veto bargaining \citep{RR:78, RR:79, Matthews:89, Cameron:00, CM:04, KKvW:21, AKK:22}.  To the best of our knowledge, none of the previous papers have allowed the agenda-setter to make the case for her proposal.  Thus the previous literature underestimates the proposer's power: Proposer's ability to engage in informative lobbying enhances her agenda-setting power (strictly when no information is suboptimal) and allows her to introduce and pass more favorable legislation.  We show that the benefit of persuasion is particularly pronounced when Proposer and Vetoer are initially expected to be quite misaligned, in which case absent persuasion the benefits of agenda-setting authority are muted.


Conceptually and methodologically, our model is closely related to the monopoly (price setting) model in which the seller chooses how much (and what) product information to provide for the buyer.\footnote{See \citet{lewis1994supplying}, \citet{che1996customer}, and \citet{anderson2006advertising} for some seminal contributions. \citet{RS:17} analyze a related problem where the buyer chooses how much information to acquire in order to influence the seller's pricing. \cite{CL:18} consider a policymaker acquiring information to increase the contributions raised from competing lobbyists.} In the price setting model, the seller can extract full surplus by providing either perfect information (if she can condition the price on the buyer's information) or no information (if she does not observe the buyer's information). In contrast, we show that, whether Proposer can observe Vetoer's information or not, it is often the case that neither perfect nor no information is optimal. The difference is driven by the partial alignment between Proposer's and Vetoer's (single-peaked) preferences, namely that, depending on his policy preferences, Vetoer may (or may not) prefer Proposer to make a more aggressive proposal; in contrast, in the price setting model, raising a price is harmful to the buyer, regardless of his valuation.


This paper proceeds as follows. \cref{sec:model} introduces the formal model, and \cref{sec:vetoer_accept} studies the optimal veto decision. \cref{sec:example} illustrates how information affects policy bargaining and \cref{sec:main} solves for the optimal experiment and proposal when Vetoer has quadratic loss.  \cref{sec:example_statics} considers specific examples and explores comparative statics. \cref{sec:linear_loss} examines how the results are changed when Vetoer has linear loss.  \cref{sec:conclude} concludes.

\section{Model}\label{sec:model} 

We build upon the canonical incomplete-information veto bargaining model. There are two players, a proposer (she) and a vetoer (he). Proposer makes a policy proposal $p\in\mathcal{R}_{+}$, and Vetoer decides whether to accept or veto it. If $p$ is accepted by Vetoer then it is implemented. If it is vetoed then the status quo policy, normalized to $0$, remains in effect.\footnote{We assume that Proposer makes a single proposal instead of a menu of options.  Unlike in a price setting model, if Vetoer has private information at the time of his decision a single proposal is generally not an optimal mechanism \citep{KKvW:21}.  However, as we illustrate in \cref{subsec:mechanisms}, under the optimal information structure a single proposal will be optimal in our baseline model.}  

Let $a\in\mathcal{R}$ denote the policy chosen, either the policy $p$ proposed by Proposer or the status quo policy $0$. Vetoer's ex post payoff is given by $v(a,\theta)=-\left(\theta-a\right)^{2}$, while Proposer's ex post payoff is given by $u(a)=-c\left(|1-a|\right)$, where $c$ is a twice continuously differentiable, strictly increasing, and convex function with $c(0)=0$. In other words, Vetoer's ideal point is $\theta$, Proposer's ideal point is $1$, Vetoer incurs utility losses that are quadratic in $|\theta-a|$, and Proposer's losses are convex in $|1-a|$.\footnote{The analysis would be unchanged if Proposer's utility were increasing in $a$ but with $1$ the maximum possible action.} As illustrated shortly, the assumption on Vetoer's quadratic preferences provides a key technical advantage for our model. 


There is imperfect information about Vetoer's ideal point $\theta$. Specifically, $\theta$ is drawn from the compact interval $\Theta:=[\underline{\theta},\overline{\theta}]$ according to the distribution $F$. For a clean analysis, we assume that $F$ has continuous and strictly positive density $f$ on $\Theta$. If $\overline{\theta}\leq 0$ then the problem becomes trivial, as Vetoer would never accept $p>0$. In what follows, we maintain the assumption that $\overline{\theta}>0$.  

Importantly, neither player observes Vetoer's ideal point $\theta$; that is, unlike in the standard incomplete-information veto bargaining, $\theta$ is \emph{not} Vetoer's private information. However, Proposer can choose to produce information about $\theta$. That Vetoer is not certain of his most-preferred policy is critical for persuasion to play a role and, as highlighted in the Introduction, is reasonable in many applications of interest. To focus on Proposer's strategic incentives, as in the recent literature on Bayesian persuasion and information design, we endow Proposer with full flexibility in her choice of information; that is, Proposer can provide any type (amount) of information. Formally, Proposer can choose any ordered set $S$ and joint distribution $\pi$ over $\Theta\times S$. Vetoer observes the realization $s\in S$ and Bayesian updates his belief. We let $F_{s}$ denote the distribution that represents Vetoer's posterior belief following the realization $s\in S$. 

We consider the following two versions of the problem. 
\begin{itemize}
    \item Persuasion first: Proposer publicly runs an experiment and makes a proposal after observing the experiment outcome. In other words, Proposer can condition his proposal $p$ on the signal realization $s$ (equivalently, Vetoer's posterior belief $F_{s}$). 
    
    \item Proposal first: This is when Proposer's proposal $p$ cannot condition on the experiment outcome. This case is relevant if Proposer makes a proposal $p$ first and then runs an experiment; in what follows, we adopt this interpretation. However, this case also arises if Proposer chooses $p$ and $F$ simultaneously; and if Proposer runs an experiment first but only Vetoer observes its outcome. 
\end{itemize}
For each case, we study weak perfect Bayesian equilibria of the given sequential-move game. This ensures that Vetoer takes the sequentially rational decision in the ``subgame'' after the proposal and outcome of the experiment have been revealed.  For convenience, we refer to weak perfect Bayesian equilibria simply as equilibria.

\section{Vetoer's Acceptance Decision}\label{sec:vetoer_accept} 

We begin by analyzing the last stage of the game in which Vetoer, with updated information about $\theta$, decides whether to accept Proposer's proposal $p$ or veto it. 

Let $V(a,G)$ denote Vetoer's expected payoff from policy $a$ when his (posterior) belief about $\theta$ can be represented by the distribution function $G$; formally, 
\begin{align*}
    V(a,G):=\mathbb{E}_{G}\left[-\left(a-\theta\right)^{2}\right]=-\int\left(a-\theta\right)^{2}dG(\theta).
\end{align*}
Vetoer is willing to accept $p$ if and only if 
\begin{align*}
    V(p,G)=-\mathbb{E}_{G}\left[\left(p-\theta\right)^{2}\right]\geq V(0,G)=-\mathbb{E}_{G}\left[\theta^{2}\right]\Leftrightarrow p\left(p-2\mathbb{E}_{G}[\theta]\right)\leq 0.
\end{align*}
This leads to the following simple but important result. 

\begin{lemma}\label{lem:quadratic_accept}
Given his posterior belief $G$, Vetoer accepts $p>0$ if and only if $p\leq 2\mathbb{E}_{G}[\theta]$. 
\end{lemma}

Since her bliss point is $1$ and she can always induce the status quo policy $0$, Proposer would never offer $p<0$. This means that \cref{lem:quadratic_accept} provides an effectively necessary and sufficient condition for Vetoer's acceptance decision. As illustrated in \cref{sec:main}, this, together with the fact that $G$ affects Vetoer's decision only through its expectation, greatly simplifies the technical analysis, thereby allowing us to study the general problem. 

If Vetoer is perfectly informed about $\theta$ then he will accept $p(\geq 0)$ if and only if $p$ is closer to $\theta$ than the status quo policy $0$, that is, $p-\theta\leq\theta$. \cref{lem:quadratic_accept} can be seen as generalizing this result to the case where there is uncertainty about $\theta$. It should be noted, however, that the result crucially depends on Vetoer's quadratic preferences; as we formally show in \cref{subsec:linear_vetoer_accept}, if Vetoer's loss function is linear then comparing $p$ to $2\mathbb{E}_{G}[\theta]$ does not suffice to determine Vetoer's decision.  


One immediate implication of \cref{lem:quadratic_accept} is that the problem becomes trivial when $\mathbb{E}[\theta]\geq 1/2$:\footnote{In what follows, we do not add subscript $F$ to $\mathbb{E}$ if the expectation is with respect to the prior distribution $F$, that is, $\mathbb{E}[\theta]=\mathbb{E}_{F}[\theta]$.} In that case Vetoer is willing to accept Proposer's ideal policy $1$ with his prior $F$. Then, it is certainly optimal for Proposer to propose her ideal policy $1$ while providing no information. In what follows, we maintain the following assumption. 
\begin{assumption}\label{assume:mean_above_half}
The unconditional expectation of $\theta$ is strictly less than $1/2$, that is, $\mathbb{E}[\theta]<1/2$.
\end{assumption}

\section{Illustration: Value of Persuasion}\label{sec:example}

This section considers a simple example in order to illustrate how Proposer can fruitfully utilize her persuasion ability. Specifically, we consider the case where Proposer's loss function is also quadratic (i.e., $c(|1-a|)=(1-a)^{2}$) and $F$ is a uniform distribution over $[\underline{\theta},1]$ for some $\underline{\theta} < 0$. 

\paragraph{No information.} Suppose Vetoer receives no additional information about his bliss point $\theta$; he only knows that $\theta$ is distributed according to $U[\underline{\theta},1]$. In this case, his expected bliss point is $\mathbb{E}[\theta]=\frac{\underline{\theta}+1}{2}$. If $\mathbb{E}[\theta]\leq 0$ then he will never accept a positive proposal (see \cref{sec:vetoer_accept}). Then, the status quo policy $0$ will be implemented, in which case Proposer earns $-c(1)=-1$. If $\mathbb{E}[\theta]>0$ then Vetoer accepts $p(>0)$ as long as $p\leq 2\mathbb{E}[\theta]$. 
In this case, Proposer's optimal proposal is $2\mathbb{E}[\theta]=\underline{\theta}+1$, resulting in a payoff of $-c(1-2\mathbb{E}[\theta])=-\underline{\theta}^{2}$. To sum up, when Vetoer obtains no information, Proposer's indirect payoff as a function of $\underline{\theta}$ is given by 
\begin{align*}
    U_{NO}(\underline{\theta}):=-\min\{\underline{\theta}^{2},1\}=\left\{\begin{array}{ll}
    -1&\text{if }\underline{\theta}<-1.\\
    -\underline{\theta}^{2}&\text{if }\underline{\theta}\in[-1,0).
    \end{array}\right.
\end{align*}

\paragraph{Full information in the proposal-first model.} Now suppose Vetoer privately observes his bliss point $\theta$; this corresponds to the standard incomplete-information veto bargaining model. In this case, a proposal $p(\geq 0)$ will be accepted if and only if $p-\theta\leq \theta-0\Leftrightarrow \theta\geq p/2$. This implies that Proposer's problem is to maximize
\begin{align*}
    -F\left(\frac{p}{2}\right)c(1)-\left(1-F\left(\frac{p}{2}\right)\right)c(1-p)=-\frac{p/2-\underline{\theta}}{1-\underline{\theta}}-\frac{1-p/2}{1-\underline{\theta}}(1-p)^{2}.
\end{align*}
The solution to this problem is $p=2/3$, and the resulting expected payoff of Proposer is 
\begin{align*}
    U_{FL,1}(\underline{\theta}):=-\frac{1/3-\underline{\theta}}{1-\underline{\theta}}-\frac{2/3}{1-\underline{\theta}}\frac{1}{9}=-\frac{11-27\underline{\theta}}{27(11-\underline{\theta})}.
\end{align*}

\begin{figure}
\begin{center}
\begin{tikzpicture}[scale=1]
		
	\draw[line width=0.5pt] (0,6)--(0,0)--(9,0)--(9,6); 
	
    \fill (0,0) node[below] {\footnotesize{$-2$}};
    \fill (9,0) node[below] {\footnotesize{$0$}};
    \fill (9,0) node[right] {\footnotesize{$\underline{\theta}$}};

    \fill (0,6) node[above] {\footnotesize{$U$}};
    
    \draw[dotted] (0,0.5)--(9,0.5);
    \draw[dotted] (0,5.5)--(9,5.5);

    \fill (0,5.5) node[left] {\footnotesize{$0$}};
    \fill (0,0.5) node[left] {\footnotesize{$-1$}};

    \draw[line width=0.5pt,blue] (0,0.5)--(4.5,0.5); 

    \fill (3,0.5) node[above] {\footnotesize{$U_{NO}$}};
    
    \draw[line width=0.5pt,blue] plot[smooth] (4.5,0.5)--(4.65,0.82778)--(4.8,1.1444)--(4.95,1.45)--(5.1,1.7444)--(5.25,2.0278)--(5.4,2.3)--(5.55,2.5611)--(5.7,2.8111)--(5.85,3.05)--(6,3.2778)--(6.15,3.4944)--(6.3,3.7)--(6.45,3.8944)--(6.6,4.0778)--(6.75,4.25)--(6.9,4.4111)--(7.05,4.5611)--(7.2,4.7)--(7.35,4.8278)--(7.5,4.9444)--(7.65,5.05)--(7.8,5.1444)--(7.95,5.2278)--(8.1,5.3)--(8.25,5.3611)--(8.4,5.4111)--(8.55,5.45)--(8.7,5.4778)--(8.85,5.4944)--(9,5.5);

    \draw[dotted] (4.5,0)--(4.5,0.5); 
    \fill (4.5,0) node[below] {\footnotesize{$-1$}};


    \draw[line width=0.5pt,brown,dashed] plot[smooth] (0,1.4877)--(0.3,1.5101)--(0.6,1.5336)--(0.9,1.5582)--(1.2,1.584)--(1.5,1.6111)--(1.8,1.6396)--(2.1,1.6696)--(2.4,1.7012)--(2.7,1.7346)--(3,1.7698)--(3.3,1.8072)--(3.6,1.8468)--(3.9,1.8889)--(4.2,1.9337)--(4.5,1.9815)--(4.8,2.0326)--(5.1,2.0873)--(5.4,2.1461)--(5.7,2.2094)--(6,2.2778)--(6.3,2.3519)--(6.6,2.4324)--(6.9,2.5202)--(7.2,2.6164)--(7.5,2.7222)--(7.8,2.8392)--(8.1,2.9691)--(8.4,3.1144)--(8.7,3.2778)--(9,3.463);

    \draw[dotted] (5.3037,0)--(5.3037,2.1265);

    \fill (3,1.7698) node[below] {\footnotesize{$U_{FL,1}$}};
    

    \draw[line width=0.5pt,black,dashdotdotted] plot[smooth] (0,1.8889)--(0.3,1.9205)--(0.6,1.9535)--(0.9,1.9881)--(1.2,2.0244)--(1.5,2.0625)--(1.8,2.1026)--(2.1,2.1447)--(2.4,2.1892)--(2.7,2.2361)--(3,2.2857)--(3.3,2.3382)--(3.6,2.3939)--(3.9,2.4531)--(4.2,2.5161)--(4.5,2.5833)--(4.8,2.6552)--(5.1,2.7321)--(5.4,2.8148)--(5.7,2.9038)--(6,3)--(6.3,3.1042)--(6.6,3.2174)--(6.9,3.3409)--(7.2,3.4762)--(7.5,3.625)--(7.8,3.7895)--(8.1,3.9722)--(8.4,4.1765)--(8.7,4.4062)--(9,4.6667);

    \draw[dotted] (5.7717,0)--(5.7717,2.9267);

    \fill (3,2.2857) node[above] {\footnotesize{$U_{FL,2}$}};

    
    \draw[line width=0.5pt,red] plot[smooth] (0,2.4753)--(0.3,2.5202)--(0.6,2.5672)--(0.9,2.6164)--(1.2,2.668)--(1.5,2.7222)--(1.8,2.7792)--(2.1,2.8392)--(2.4,2.9024)--(2.7,2.9691)--(3,3.0397)--(3.3,3.1144)--(3.6,3.1936)--(3.9,3.2778)--(4.2,3.3674)--(4.5,3.463)--(4.8,3.5651)--(5.1,3.6746)--(5.4,3.7922)--(5.7,3.9188)--(6,4.0556)--(6.3,4.2037)--(6.6,4.3647)--(6.9,4.5404)--(7.2,4.7328)--(7.5,4.9444)--(7.8,5.1444)--(8.1,5.3)--(8.4,5.4111)--(8.7,5.4778)--(9,5.5);

    \fill (3,3.0397) node[above] {\footnotesize{$U_{BI}$}};

    \draw[dotted] (7.5,0)--(7.5,89/18); 
    \fill (7.5,0) node[below] {\footnotesize{$-\frac{1}{3}$}};
    
	\end{tikzpicture}
\end{center}\caption{\label{fig:example_compare}This figure depicts Proposer's indirect expected payoffs as a function of $\underline{\theta}$ under four different information regimes. 
}
\end{figure}
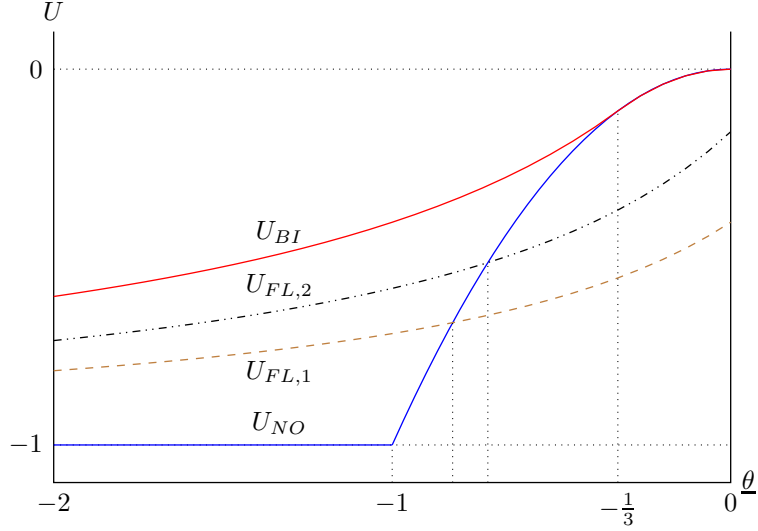

As depicted in \cref{fig:example_compare}, $U_{NO}$ crosses $U_{FL,1}$ once from below. 
With no information, 
Vetoer believes that his ideal action may be very low and so is unwilling to accept a policy much above the status quo. Full information allows Proposer to avoid this problem as now she can target only high Vetoer types. However, it comes at the cost of yielding some information rent to Vetoer. The fact that the cost associated with no information declines fast as $\underline{\theta}$ rises illustrates the particular single-crossing pattern. 

\paragraph{Full information in the persuasion-first model.} What if Proposer also observes Vetoer's bliss point $\theta$, that is, $\theta$ becomes public before Proposer chooses a proposal? In that case, Proposer offers $0$ if $\theta\leq 0$, $2\theta$ if $\theta\in(0,1/2)$, and $1$ if $\theta\geq 1/2$. The resulting payoff of Proposer is 
\begin{align*}
    U_{FL,2}(\underline{\theta})&:=-\int_{\underline{\theta}}^{0}c(1)dF(\theta)-\int_{0}^{1/2}c(1-2\theta)dF(\theta)-\int_{1/2}^{1}c(0)dF(\theta)\\
    &=-\int_{\underline{\theta}}^{0}1\frac{d\theta}{1-\underline{\theta}}-\int_{0}^{1/2}\left(1-2\theta\right)^{2}\frac{d\theta}{1-\underline{\theta}}=-\frac{1}{1-\underline{\theta}}\left(\frac{1}{6}-\underline{\theta}\right).
\end{align*}

As shown in \cref{fig:example_compare}, $U_{FL,2}$ is uniformly above $U_{FL,1}$. This holds simply because Proposer has more flexibility under persuasion first than under proposal first; 
she can always mimic her proposal-first strategy under persuasion first by choosing to not condition her proposal on the experiment outcome, but not vice versa. It is more intriguing that $U_{FL,2}<U_{NO}$ when $\underline{\theta}$ is relatively close to $0$, at which point the resulting action with no information is close to Proposer's bliss point. Note that the current full-information persuasion-first case corresponds to the perfect price discrimination in the price setting model (which yields the first-best outcome to the seller). The comparison result is driven by the following two facts: (i) Vetoer (still) obtains information rents when $\theta\in(1/2,1]$, and (ii) Proposer is risk averse and full information results in a random outcome (depending on $\theta$). 

\paragraph{Optimal binary signal.} Consider a simple binary signal that produces $0$ if $\theta<s_{\ast}$ and $1$ if $\theta\geq s_{\ast}$ for some $s_{\ast}\leq[\max\{\underline{\theta},-1\},0]$. If the signal realization is $0$ then Vetoer's posterior belief is uniform over $[\underline{\theta},s_{\ast})$. In this case, $\mathbb{E}[\theta|0]=\frac{\underline{\theta}+s_{\ast}}{2}<0$, so Proposer has no choice but to induce the status quo policy $0$. If the realization is $1$ then Vetoer's posterior belief is uniform over $[s_{\ast},1]$. In this case, $\mathbb{E}[\theta|1]\geq 0$, so Proposer's optimal proposal is $2\mathbb{E}[\theta|1]=s_{\ast}+1$. This implies that Proposer chooses $s_{\ast}$ to maximize: 
\begin{align*}
        -F(s_{\ast})c(1)-(1-F(s_{\ast}))c\left(1-2\mathbb{E}[\theta|1]\right)=-\frac{s_{\ast}-\underline{\theta}}{1-\underline{\theta}}-\frac{(1-s_{\ast})s_{\ast}^{2}}{1-\underline{\theta}}. 
\end{align*}
The solution to this problem is $s_{\ast}=\max\{\underline{\theta},-1/3\}$, and the resulting indirect utility of Proposer is 
\begin{align*}
        U_{BI}(\underline{\theta}):=\left\{\begin{array}{ll}
        -\frac{1}{1-\underline{\theta}}\left(-\frac{5}{27}-\underline{\theta}\right)&\text{if }\underline{\theta}<-\frac{1}{3}\\
        -\underline{\theta}^{2}&\text{if }\underline{\theta}\geq-\frac{1}{3}.
        \end{array}\right.
\end{align*}
Notice that our derivation is based on the persuasion-first interpretation. It is straightforward, however, that Proposer can obtain the same outcome under proposal first; it suffices that she employs the same binary experiment and proposes $2\mathbb{E}[\theta|1]$. Note that this equivalence holds because---and whenever---Proposer makes only one strictly positive proposal under persuasion first. 

\paragraph{Comparison.}  As depicted in \cref{fig:example_compare}, this optimal binary information uniformly dominates all the above cases. This is because a binary signal offers a simple but powerful way to avoid the problems of both no information and full information. Unlike no information, it enables Proposer to exclude Vetoer types below $s_{\ast}$ and so be aggressive in her proposal; binary information strictly dominates no information unless $\underline{\theta}$ is sufficiently large. Meanwhile, unlike full information, it leaves no rent to Vetoer and induces a single positive policy. Can Proposer do better than $U_{BI}$ by choosing a more sophisticated experiment? Our subsequent general analysis shows that the answer is no, that is, the optimal binary signal is an optimal signal among all possible ones. 

\section{Main Characterization}\label{sec:main}

This section provides a full characterization for our general model. 

\subsection{Persuasion-First Model}\label{subsec:persuade_first}

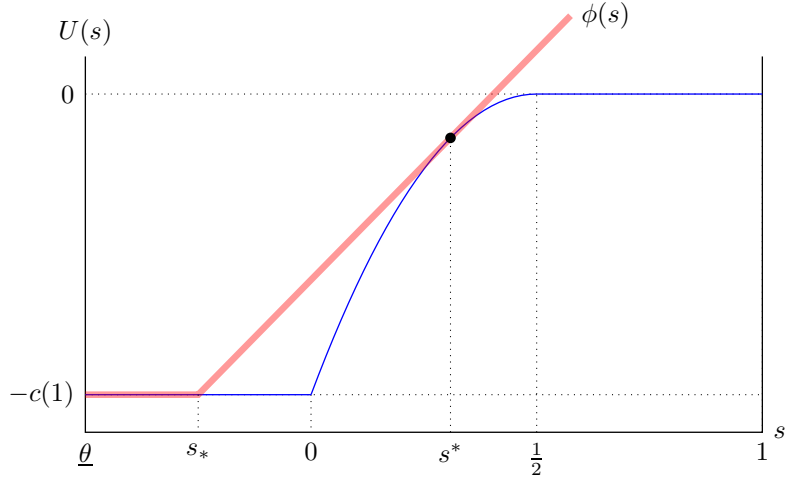
\begin{figure}
\begin{center}
\begin{tikzpicture}[scale=1]
		
	\draw[line width=0.5pt] (-3,5)--(-3,0)--(6,0)--(6,5);  
	
    \fill (-3,0) node[below] {\footnotesize{$\underline{\theta}$}};
    \fill (0,0) node[below] {\footnotesize{$0$}};
    \fill (3,0) node[below] {\footnotesize{$\frac{1}{2}$}};
    \fill (6,0) node[below] {\footnotesize{$1$}};

    \draw[dotted] (0,0.5)--(0,0);

    \fill (6,0) node[right] {\footnotesize{$s$}};
    \fill (-3,5) node[above] {\footnotesize{$U(s)$}};

    \draw[dotted] (-3,0.5)--(6,0.5); 
    \fill (-3,0.5) node[left] {\footnotesize{$-c(1)$}};

    \draw[dotted] (3,0)--(3,4.5); 
    \draw[dotted] (6,0)--(6,4.5); 

    \draw[dotted] (-3,4.5)--(6,4.5); 
    \fill (-3,4.5) node[left] {\footnotesize{$0$}};

    \draw[line width=0.5pt,blue] plot[smooth] (0,0.5)--(0.1,0.76222)--(0.2,1.0156)--(0.3,1.26)--(0.4,1.4956)--(0.5,1.7222)--(0.6,1.94)--(0.7,2.1489)--(0.8,2.3489)--(0.9,2.54)--(1,2.7222)--(1.1,2.8956)--(1.2,3.06)--(1.3,3.2156)--(1.4,3.3622)--(1.5,3.5)--(1.6,3.6289)--(1.7,3.7489)--(1.8,3.86)--(1.9,3.9622)--(2,4.0556)--(2.1,4.14)--(2.2,4.2156)--(2.3,4.2822)--(2.4,4.34)--(2.5,4.3889)--(2.6,4.4289)--(2.7,4.46)--(2.8,4.4822)--(2.9,4.4956)--(3,4.5);

    \draw[line width=0.5pt,blue] (-3,0.5)--(0,0.5); 
    \draw[line width=0.5pt,blue] (3,4.5)--(6,4.5); 

    \draw[line width=2.5pt,red, draw opacity=0.4] (-3,0.5)--(-1.5,0.5)--(3.45,5.542); 

    \draw[dotted] (-1.5,0.5)--(-1.5,0); 
    \fill (-1.5,0) node[below] {\footnotesize{$s_{\ast}$}};

   \fill (3.45,5.542) node[right] {\footnotesize{$\phi(s)$}};

    \draw[dotted] (1.8541,0)--(1.8541,3.9164);
    \fill[black] (1.8541,3.9164) circle (2pt); 

    \fill (1.8541,0) node[below] {\footnotesize{$s^{\ast}$}};
    

	\end{tikzpicture}
\end{center}\caption{\label{fig:indirect_util_DM}This figure depicts Proposer's indirect utility function $U(s)$ (blue solid) and a convex function $\phi(s)$ (red translucent) that can be used to apply \citeauthor{DM:19}'s results. In this figure, $c$ is quadratic.}
\end{figure}

We first consider the persuasion-first model. Recall that given the realization $s\in S$ (or Vetoer's posterior belief $F_{s}$), Vetoer is willing to accept any $p\leq 2\mathbb{E}[\theta|s]=2\mathbb{E}_{F_{s}}[\theta]$ (\cref{lem:quadratic_accept}). Combining this with the fact that Proposer can always choose to induce the status quo policy $0$ and her bliss point is $1$, it follows that Proposer's optimal proposal is given by 
\begin{align*}
    p^{\ast}(s)=\left\{\begin{array}{ll}
    0&\text{if }\mathbb{E}[\theta|s]\leq 0\\
    2\mathbb{E}[\theta|s]&\text{if }\mathbb{E}[\theta|s]\in(0,1/2)\\
    1&\text{if }\mathbb{E}[\theta|s]\geq 1/2.
    \end{array}\right.
\end{align*}
The corresponding indirect utility of Proposer is  
\begin{align*}
    U(s):=u(p^{\ast}(s))=\left\{\begin{array}{ll}
    -c(1)&\text{if }\mathbb{E}[\theta|s]\leq 0\\
    -c(1-2\mathbb{E}[\theta|s])&\text{if }\mathbb{E}[\theta|s]\in(0,1/2)\\
    0&\text{if }\mathbb{E}[\theta|s]\geq 1/2.
    \end{array}\right.
\end{align*}
See the blue solid curve in \cref{fig:indirect_util_DM} for the representative shape of $U$. 

\paragraph{Reformulation in terms of the distribution of posterior means.} Crucially, Proposer's indirect utility depends only on the posterior mean $\mathbb{E}[\theta|s]$, not on other aspects of $F_{s}$; without loss of generality, we use $s$ to denote the mean of the posterior distribution following realization $s$. This implies that we can focus on the distribution over \emph{posterior means}, not the distribution over \emph{posterior distributions}. Then, we can utilize a powerful fact that given $F$ there exists an experiment that produces a distribution $G$ of posterior means if and only if $F$ is a mean-preserving spread of $G$.\footnote{See, e.g., \citet{blackwell1953equivalent,gentzkow2016rothschild,kolotilin2018optimal}. Intuitively, $G=F$ if the experiment is perfectly informative, while $G$ is degenerate at $\mathbb{E}[\theta]$ if the experiment carries no information. Any intermediate level of information produces a distribution between these two extremes.} This allows us to write Proposer's optimal persuasion problem as
\begin{align}\label{eq:persuade_first_problem}
    \max_{G\in MPC(F)}\int U(s)dG(s),
\end{align}
where $MPC(F)$ denotes the set of all mean-preserving contractions of $F$. 

To solve \eqref{eq:persuade_first_problem}, we apply a recent theoretical development by \citet{DM:19}.\footnote{No elementary method can be used to analyze \eqref{eq:persuade_first_problem}, because it is an infinite-dimensional programming problem with a complex constraint ($G\in MPC(F)$).} They provide a simple geometric verification tool for a class of Bayesian persuasion problems that can be written in the form of \eqref{eq:persuade_first_problem}. To apply their technique, we first plot $U(s)$; see the blue solid curve in \cref{fig:indirect_util_DM}. Then, we consider a smallest convex function $\phi$ that uniformly stays above $U$ and identify the set of values such that $\phi(s)=U(s)$; in \cref{fig:indirect_util_DM}, the red translucent curve represents such a convex function, and $\{s:\phi(s)=U(s)\}=[\underline{\theta},s_{\ast}]\cup\{s^{\ast}\}$. The remaining task is to see whether there is a distribution $G$ such that $supp(G)\subseteq\{s:\phi(s)=U(s)\}$ and $\int\phi(s)dF(s)=\int\phi(s)dG(s)$. In our problem, due to the piece-wise linear structure of $\phi$, this last requirement reduces to
\begin{align}
\label{e:star}
    s^{\ast}=\mathbb{E}[\theta|\theta\geq s_{\ast}]=\int_{s_{\ast}}^{\overline{\theta}}\theta \frac{dF(\theta)}{1-F(s_{\ast})}. 
\end{align}
Note that there is a one-to-one mapping between the cutoff $s_*$ and the expectation $s^*$. Applying this procedure to our problem yields the following result. 

\begin{proposition}\label{prop:quadratic_persuade_first}
    In the persuasion-first model, no information is optimal to Proposer if and only if
    \begin{align}
        U(\underline{\theta})\leq U(\mathbb{E}[\theta])+U^{\prime}(\mathbb{E}[\theta])(\underline{\theta}-\mathbb{E}[\theta]).\label{eq:no_info_opt}
    \end{align}
    If this condition fails then the binary signal that reveals only whether $\theta\geq s_{\ast}$ or not is optimal to Proposer,  
    where $s_{\ast}(\leq 0)$ is the value such that for all $s\in[s_{\ast},\overline{\theta}]$, 
    \begin{align}
        U(s)\leq U(s_{\ast})+\frac{U(\mathbb{E}[\theta|\theta\geq s_{\ast}])-U(s_{\ast})}{\mathbb{E}[\theta|\theta\geq s_{\ast}]-s_{\ast}}(s-s_{\ast}).\label{eq:some_info_opt}
    \end{align}
\end{proposition}

Most notable in \cref{prop:quadratic_persuade_first} is that the proposer-optimal outcome can be achieved with either no information or a simple binary experiment. This emerges because Proposer is risk averse over policy outcomes and, as shown above, the implemented policy $p^{\ast}(s)$ is linear in the posterior mean $\mathbb{E}[\theta|s]$ over the relevant region $[0,1]$. To see this clearly, suppose there are two signal realizations that lead to two different positive policies. If Proposer pools the two realizations into one then she can achieve the same expected policy with no risk, which is always beneficial to (risk-averse) Proposer. 
Meanwhile, Proposer seeks to maximize the expected mean after a positive message.  This explains why the optimal binary signal takes a simple cutoff structure.\footnote{The optimality of an experiment with at most two messages is reminiscent of the classical result in \cite{Matthews:89} on size-two equilibria with cheap talk and veto threats.  In addition to the very different environments---\cite{Matthews:89} considers cheap talk by a privately informed veto player---the logic is quite different.  The key to a binary cheap talk message is that Proposer always offers her bliss point if not receiving a veto threat, whereas a binary message emerges in our setting due to Proposer's risk aversion over policy outcomes.}  

 
Since $U(s)=u(p^{\ast}(s))$ and $p^{\ast}=2\mathbb{E}[\theta|s]$ whenever $\mathbb{E}[\theta|s]\in(0,1/2)$, \eqref{eq:no_info_opt} can be rewritten as 
\[
2u^{\prime}(2\mathbb{E}[\theta])(\mathbb{E}[\theta]-\underline{\theta}) \leq u(2\mathbb{E}[\theta])-u(0).
\]
To understand this condition, suppose $\underline{\theta}<0$ and $\mathbb{E}[\theta]>0$,\footnote{If $\underline{\theta} \geq 0$ then the experiment only creates uncertainty in the resulting policy, without changing the expected policy. Therefore, the experiment is never beneficial to Proposer. If $\mathbb{E}[\theta]\leq 0$ then no information is obviously not optimal.} and compare no information with an experiment which reveals if $\theta$ is in a small neighborhood of $\underline{\theta}$ or not (i.e., whether $\theta<\underline{\theta}+\varepsilon$ or not for $\varepsilon$ sufficiently small). This experiment raises the induced positive policy (from $\mathbb{E}[\theta]$ to $\mathbb{E}[\theta|\theta\geq\underline{\theta}+\varepsilon]$), but now results in policy $0$, instead of $2\mathbb{E}[\theta]$, when $\theta<\underline{\theta}+\varepsilon$. The left-hand side of the above inequality captures the former marginal benefit (of revealing whether $\theta<\underline{\theta}+\varepsilon$), while the right-hand side represents the corresponding marginal cost. In other words, \eqref{eq:no_info_opt} coincides with the condition under which Proposer has no incentive to reveal $\theta$ even when it is arbitrarily close to $\underline{\theta}$. Note that it follows immediately that if $\underline{\theta}$ is sufficiently negative relative to $\mathbb{E}[\theta]$, then Proposer necessarily benefits from providing some information. This is because when $\theta<\underline{\theta}+\varepsilon$ is revealed the resulting action is $0$, so when $\underline{\theta}$ is sufficiently low providing some information increases the expected action significantly.  


When no information is suboptimal, the optimal binary experiment is given by revealing whether or not $\theta \geq s_*$, defined by \eqref{eq:some_info_opt}. Note that when $u$ is differentiable (i.e., no kink at Proposer's bliss point), this inequality can only be satisfied for all $s$ if
\[
U'(s_*)=\frac{U(\mathbb{E}[\theta|\theta\geq s_{\ast}])-U(s_{\ast})}{\mathbb{E}[\theta|\theta\geq s_{\ast}]}
\]
or equivalently 
\[
2u'(2E[\theta|\theta \geq s_*])(E[\theta|\theta \geq s_*]-s_*)=u(2E[\theta|\theta \geq s_*])-u(0).
\]
This condition determines the $s_*$ for which \eqref{eq:no_info_opt} holds with equality when $\underline{\theta}=s_*$.  In particular, this implies that $s_* \leq 0$: $s_*=0$ maximizes the expected action, but a risk averse Proposer has an added incentive to reduce vetoes.  When $u$ has a kink at $a=1$ it is also possible to get a ``corner'' solution in which $s^*=1/2$ and so the resulting proposal is either $0$ or $1$; see \cref{subsec:linear_proposer}.

Finally, we note that when \eqref{eq:no_info_opt} fails, there exist multiple optimal experiments; any experiment that reveals whether $\theta\geq s_{\ast}$ or not, but provides no further information above $s_{\ast}$ is optimal. For example, it is optimal to fully reveal $\theta$ below $s_{\ast}$ but provide no other information. This multiplicity arises because for $\theta<s_{\ast}$, the status quo policy is implemented and so the fine detail of the experiment is inconsequential for the outcome. 

\begin{figure}
\begin{center}
\begin{tikzpicture}[scale=0.85]
		
	\draw[line width=0.5pt] (-3,5)--(-3,0)--(3,0)--(3,5);  
	
    \fill (-3,0) node[below] {\footnotesize{$\underline{\theta}$}};
    \fill (0,0) node[below] {\footnotesize{$0$}};
    \fill (3,0) node[below] {\footnotesize{$\overline{\theta}=\frac{1}{2}$}};

    \draw[dotted] (0,0.5)--(0,0);

    \fill (3,0) node[right] {\footnotesize{$s$}};
    \fill (-3,5) node[above] {\footnotesize{$U(s)$}};

    \draw[dotted] (-3,0.5)--(3,0.5); 
    \fill (-3,0.5) node[left] {\footnotesize{$-c(1)$}};


    \draw[dotted] (-3,4.5)--(3,4.5); 
    \fill (-3,4.5) node[left] {\footnotesize{$0$}};

    \draw[line width=0.5pt,blue] (-3,0.5)--(0,0.5);

    \draw[line width=0.5pt,blue] plot[smooth] (0,0.5)--(0.1,0.56723)--(0.2,0.63563)--(0.3,0.70527)--(0.4,0.7762)--(0.5,0.84852)--(0.6,0.92229)--(0.7,0.99762)--(0.8,1.0746)--(0.9,1.1534)--(1,1.234)--(1.1,1.3167)--(1.2,1.4016)--(1.3,1.4889)--(1.4,1.5788)--(1.5,1.6716)--(1.6,1.7675)--(1.7,1.8669)--(1.8,1.9702)--(1.9,2.0779)--(2,2.1906)--(2.1,2.3091)--(2.2,2.4344)--(2.3,2.5678)--(2.4,2.7111)--(2.4207,2.7423)--(2.4414,2.7739)--(2.4621,2.8062)--(2.4828,2.8391)--(2.5034,2.8726)--(2.5241,2.9069)--(2.5448,2.9419)--(2.5655,2.9778)--(2.5862,3.0144)--(2.6069,3.0521)--(2.6276,3.0907)--(2.6483,3.1304)--(2.669,3.1713)--(2.6897,3.2135)--(2.7103,3.2571)--(2.731,3.3023)--(2.7517,3.3493)--(2.7724,3.3983)--(2.7931,3.4495)--(2.8138,3.5035)--(2.8345,3.5604)--(2.8552,3.6211)--(2.8759,3.6863)--(2.8966,3.7572)--(2.9172,3.8356)--(2.9379,3.9246)--(2.9586,4.0302)--(2.9793,4.1678)--(3,4.5);
    \draw[line width=0.5pt,blue] (-3,0.5)--(0,0.5);

    \draw[line width=2.5pt,red, draw opacity=0.4]  (-3,0.5+0.03)--(0,0.5+0.03); 

    \draw[line width=2.5pt,red, draw opacity=0.4] plot[smooth] (0,0.5+0.05)--(0.1,0.56723+0.05)--(0.2,0.63563+0.05)--(0.3,0.70527+0.05)--(0.4,0.7762+0.05)--(0.5,0.84852+0.05)--(0.6,0.92229+0.05)--(0.7,0.99762+0.05)--(0.8,1.0746+0.05)--(0.9,1.1534+0.05)--(1,1.234+0.05)--(1.1,1.3167+0.05)--(1.2,1.4016+0.05)--(1.3,1.4889+0.05)--(1.4,1.5788+0.05)--(1.5,1.6716+0.05)--(1.6,1.7675+0.05)--(1.7,1.8669+0.05)--(1.8,1.9702+0.05)--(1.9,2.0779+0.05)--(2,2.1906+0.05)--(2.1,2.3091+0.05)--(2.2,2.4344+0.05)--(2.3,2.5678+0.05)--(2.4,2.7111+0.05)--(2.4207,2.7423+0.05)--(2.4414,2.7739+0.05)--(2.4621,2.8062+0.05)--(2.4828,2.8391+0.05)--(2.5034,2.8726+0.05)--(2.5241,2.9069+0.05)--(2.5448,2.9419+0.05)--(2.5655,2.9778+0.05)--(2.5862,3.0144+0.05)--(2.6069,3.0521+0.05)--(2.6276,3.0907+0.05)--(2.6483,3.1304+0.05)--(2.669,3.1713+0.05)--(2.6897,3.2135+0.05)--(2.7103,3.2571+0.05)--(2.731,3.3023+0.05)--(2.7517,3.3493+0.05)--(2.7724,3.3983+0.05)--(2.7931,3.4495+0.05)--(2.8138,3.5035+0.05)--(2.8345,3.5604+0.05)--(2.8552,3.6211+0.05)--(2.8759,3.6863+0.05)--(2.8966,3.7572+0.05)--(2.9172,3.8356+0.05)--(2.9379,3.9246+0.05)--(2.9586,4.0302+0.05)--(2.9793,4.1678+0.05)--(3,4.5+0.05);




    

    \begin{scope}[xshift=8cm]
    
    \draw[line width=0.5pt] (-3,5)--(-3,0)--(4.5,0)--(4.5,5);  
	
    \fill (-3,0) node[below] {\footnotesize{$\underline{\theta}$}};
    \fill (0,0) node[below] {\footnotesize{$0$}};
    \fill (3,0) node[below] {\footnotesize{$s^{\ast}=\frac{1}{2}$}};
    \fill (4.5,0) node[below] {\footnotesize{$\overline{\theta}$}};

    \draw[dotted] (0,0.5)--(0,0);

    \fill (4.5,0) node[right] {\footnotesize{$s$}};
    \fill (-3,5) node[above] {\footnotesize{$U(s)$}};

    \draw[dotted] (-3,0.5)--(4.5,0.5); 
    \fill (-3,0.5) node[left] {\footnotesize{$-c(1)$}};

    \draw[dotted] (3,0)--(3,4.5); 

    \draw[dotted] (-3,4.5)--(4.5,4.5); 
    \fill (-3,4.5) node[left] {\footnotesize{$0$}};

    \draw[line width=0.5pt,blue] plot[smooth] (0,0.5)--(0.1,0.56723)--(0.2,0.63563)--(0.3,0.70527)--(0.4,0.7762)--(0.5,0.84852)--(0.6,0.92229)--(0.7,0.99762)--(0.8,1.0746)--(0.9,1.1534)--(1,1.234)--(1.1,1.3167)--(1.2,1.4016)--(1.3,1.4889)--(1.4,1.5788)--(1.5,1.6716)--(1.6,1.7675)--(1.7,1.8669)--(1.8,1.9702)--(1.9,2.0779)--(2,2.1906)--(2.1,2.3091)--(2.2,2.4344)--(2.3,2.5678)--(2.4,2.7111)--(2.4207,2.7423)--(2.4414,2.7739)--(2.4621,2.8062)--(2.4828,2.8391)--(2.5034,2.8726)--(2.5241,2.9069)--(2.5448,2.9419)--(2.5655,2.9778)--(2.5862,3.0144)--(2.6069,3.0521)--(2.6276,3.0907)--(2.6483,3.1304)--(2.669,3.1713)--(2.6897,3.2135)--(2.7103,3.2571)--(2.731,3.3023)--(2.7517,3.3493)--(2.7724,3.3983)--(2.7931,3.4495)--(2.8138,3.5035)--(2.8345,3.5604)--(2.8552,3.6211)--(2.8759,3.6863)--(2.8966,3.7572)--(2.9172,3.8356)--(2.9379,3.9246)--(2.9586,4.0302)--(2.9793,4.1678)--(3,4.5);
    
    \draw[line width=0.5pt,blue] (-3,0.5)--(0,0.5);

    \draw[line width=0.5pt,blue] (3,4.5)--(4.5,4.5);

    \draw[line width=2.5pt,red, draw opacity=0.4]  (-3,0.5)--(-1.5,0.5); 

    \draw[line width=2.5pt,red, draw opacity=0.4]  (-1.5,0.5)--(3,4.5)--(4,4.5+8/9);

    \draw[dotted] (-1.5,0.5)--(-1.5,0); 
    \fill (-1.5,0) node[below] {\footnotesize{$s_{\ast}$}};

   \fill[black] (3,4.5) circle (2pt);

    \end{scope}

	\end{tikzpicture}
\end{center}\caption{\label{fig:concave_loss}Both panels depict Proposer's indirect utility function $U(s)$ (blue solid) and the smallest convex function $\phi(s)$ above $U$ when $u$ is convex ($c$ concave). They differ in that $\overline{\theta}\leq 1/2$ in the left panel, while $\overline{\theta}>1/2$ in the right panel.}
\end{figure}
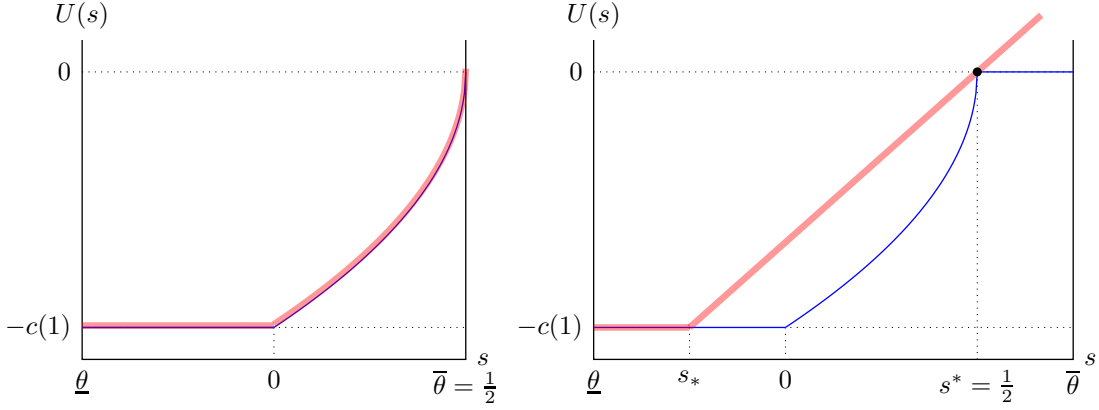

\paragraph{Optimality of full information.} As is well known, full information is optimal if and only if $U$ is convex over $[\underline{\theta},\overline{\theta}]$. Intuitively, the agent is ``risk loving'' and so wishes to disperse the distribution $G$ of posterior means as much as possible. In $MPC(F)$, the most dispersed distribution is $F$ itself. In our problem, this condition holds if and only if, as depicted in the left panel of \cref{fig:concave_loss}, $\overline{\theta}\leq 1/2$ and $u$ is \emph{convex} ($c$ is \emph{concave}) over $[0,2\overline{\theta}]$; the former ensures that $U$ does not have the second flat region (above $1/2$), while the latter is necessary and sufficient for $U$ to be convex over $[0,\overline{\theta}]$. Note that if $\overline{\theta}>1/2$ then, even if $u$ is convex, full information is suboptimal: Proposer will never reveal that $\theta>1/2$ as she would prefer to pool such vetoers with some below $1/2$, securing her ideal action of $a=1$ more often.  See the right panel of \cref{fig:concave_loss}.

\subsection{Proposal-First Model}\label{subsec:proposal_first}

We now turn to the proposal-first model where Proposer cannot tailor her proposal to the outcome of the experiment. In this case, one can first find the optimal signal for each proposal $p$ and then optimize over $p$; this is the approach we take for the linear Vetoer case in \cref{subsec:linear_propose_first}. For our main quadratic case, we take a different (and arguably more elegant) approach. By definition, Proposer has more flexibility under persuasion-first than under proposal-first. 
Therefore, the optimal persuasion-first outcome provides an upper bound for her expected payoff in the proposal-first model. We show that the optimal persuasion-first outcome in \cref{prop:quadratic_persuade_first} can be implemented in the proposal-first model. 



\begin{proposition}\label{prop:proposal_first} 
    The optimal experiments in \cref{prop:quadratic_persuade_first} remain optimal in the proposal-first model. Proposer optimally proposes $2\mathbb{E}[\theta]$ if \eqref{eq:no_info_opt} holds (i.e., no information is optimal) and $2\mathbb{E}[\theta|\theta\geq s_{\ast}]$ otherwise. 
\end{proposition}

\cref{prop:proposal_first} holds because the proposer-optimal in the persuasion-first model involves only one positive proposal; Proposer can achieve the same payoff as under proposal-first by making that positive proposal and accepting a veto otherwise. The optimality of a single positive proposal holds whenever $u$ is concave. If $u$ is strictly convex and $\overline{\theta}\in(0,1/2)$, however, the optimal signal under persuasion-first should fully reveal all $\theta$'s above $0$. Such an outcome clearly cannot be implemented in the proposal-first model. \cref{prop:proposal_first} also relies on the assumption on Vetoer's quadratic preferences. As we formally show in \cref{sec:linear_loss}, if Vetoer incurs linear losses then Proposer can often do strictly better under persuasion-first than under proposal-first. 

\subsection{Discussion of Assumptions}\label{subsec:mechanisms} 

A novel theoretical implication of \cref{prop:proposal_first} is that a single take-it-or-leave-it offer is optimal even in the class of all mechanisms in our model. In the persuasion-first model, it is natural because Proposer and Vetoer are symmetrically informed when the proposal is made and, therefore, Proposer can always offer the highest policy Vetoer would accept. In the proposal-first model, however, Vetoer acquires some private information prior to his decision, so in principle Proposer could benefit from offering a menu of alternatives \citep{KKvW:21}. \cref{prop:proposal_first} shows that (with quadratic loss) such mechanism design is of no use to Proposer in our model where she engages in optimal information design; Proposer can obtain the optimal persuasion-first outcome even with a single take-it-or-leave-it offer. 

Furthermore, achieving the persuasion-first outcome does not depend on Proposer being able to commit not to revise her proposal in the event of a veto.  While in general Proposer may want to do so \citep{Cameron:00, AKK:22}, under the optimal experiment vetoes occur only after Vetoer's bliss point is revealed to be below the status quo, in which case no mutually preferred policy exists. As such, Proposer commitment is immaterial. 

However, Proposer's ex ante commitment to an experiment (and truthfully revealing its outcome) is crucial in implementing her optimal outcome. Specifically, Proposer cannot necessarily obtain her optimal outcome if she privately observes Vetoer's ideal point $\theta$ and communicates with Vetoer through cheap talk \`{a} la \citet{CS:82} or voluntary disclosure \`{a} la \citet{grossman1981informational} and \citet{milgrom1981good}. Cheap talk cannot be effective because Proposer's preferences are independent of $\theta$. Voluntary disclosure, in which Proposer can disclose any set containing $\theta$ but cannot lie, can implement Proposer's optimal outcome if and only if $s^{\ast}=\mathbb{E}[\theta|s\geq s_{\ast}]=1/2$, so that the induced policy is either the status quo $0$ or Proposer's ideal one $1$; this is because Proposer has an incentive to reveal $\theta\geq 1/2$, which is necessary and sufficient to trigger familiar unraveling, if and only if $s^{\ast}<1/2$.\footnote{This discussion assumes Proposer privately knows $\theta$ at the beginning of the game.  If the players are symmetrically informed at the beginning of the game, and Proposer only privately learns $\theta$ after committing to a proposal, then as in \citet{Titova:22}, there is an equilibrium in which Proposer only reveals whether or not $\theta \geq s_*$ in the disclosure stage. In this way proposal-first can outperform disclosure-first, provided the proposal is made prior to Proposer becoming informed.}

\section{Examples and Comparative Statics}\label{sec:example_statics}

This section applies the above characterization to two leading examples where Proposer's loss function $c$ is either linear or quadratic. It also studies how Proposer's optimal strategy and the resulting outcome depend on the characteristics of the underlying veto bargaining problem. 



\subsection{Linear Proposer}\label{subsec:linear_proposer}
\begin{figure}
\centering
\begin{subfigure}{0.325\textwidth}
    \begin{tikzpicture}[xscale=4/1,yscale=4/1.2]
		
	\draw[line width=0.5pt] (0,0.1)--(0,-1.1)--(1,-1.1)--(1,0.1);  
	
    \fill (0,-1.1) node[below] {\footnotesize{$\underline{\theta}=0$}};
    \fill (0.5,-1.1) node[below] {\footnotesize{$\frac{1}{2}$}};
    \fill (1,-1.1) node[below] {\footnotesize{$\overline{\theta}$}};
    \fill (0,0.1) node[above] {\footnotesize{$U(s)$}};
    \fill (0,0) node[left] {\footnotesize{$0$}};
    \fill (0,-1) node[left] {\footnotesize{$-c(1)$}};
    \fill (1,-1.1) node[right] {\footnotesize{$s$}};

    \draw[dotted] (0,0)--(.5,0);
    \draw[dotted] (0,-1)--(1,-1);
    \draw[dotted] (0.5,0)--(0.5,-1.1);
    \draw[line width=0.5pt,blue] (0,-1)--(0.5,0)--(1,0);
    \draw[line width=2.5pt,red, draw opacity=0.4] (0,-1)--(0.5,0)--(1,0);

	\end{tikzpicture}
\end{subfigure}
\begin{subfigure}{0.325\textwidth}
    \begin{tikzpicture}[xscale=4/1.5,yscale=4/1.2]
		
	\draw[line width=0.5pt] (-0.5,0.1)--(-0.5,-1.1)--(1,-1.1)--(1,0.1);  
	
    \fill (-0.5,-1.1) node[below] {\footnotesize{$\underline{\theta}$}};
    \fill (0,-1.1) node[below] {\footnotesize{$0$}};
    \fill (-0.25,-1.1) node[below] {\footnotesize{$s_{\ast}$}};
    \fill (0.5,-1.1) node[below] {\footnotesize{$\frac{1}{2}$}};
    \fill (1,-1.1) node[below] {\footnotesize{$\overline{\theta}$}};
    \fill (-0.5,0.1) node[above] {\footnotesize{$U(s)$}};
    \fill (-0.5,0) node[left] {\footnotesize{$0$}};
    \fill (-0.5,-1) node[left] {\footnotesize{$-c(1)$}};
    \fill (1,-1.1) node[right] {\footnotesize{$s$}};
    
    \draw[dotted] (-0.5,0)--(1,0);
    \draw[dotted] (-0.5,-1)--(1,-1);
    \draw[dotted] (-0.25,-1)--(-0.25,-1.1);
    \draw[dotted] (0,-1)--(0,-1.1);
    \draw[dotted] (0.5,0)--(0.5,-1.1);
    \draw[line width=0.5pt,blue] (-0.5,-1)--(0,-1)--(0.5,0)--(1,0);
    \draw[line width=2.5pt,red, draw opacity=0.4] (-0.5,-1)--(-0.25,-1)--(0.575,0.1);

	\end{tikzpicture}
\end{subfigure}
\begin{subfigure}{0.325\textwidth}
    \begin{tikzpicture}[xscale=4/1.5,yscale=4/1.2]
		
	\draw[line width=0.5pt] (-0.5,0.1)--(-0.5,-1.1)--(1,-1.1)--(1,0.1);  
	
    \fill (-0.5,-1.1) node[below] {\footnotesize{$\underline{\theta}$}};
    \fill (0,-1.1) node[below] {\footnotesize{$0$}};
    \fill (0.25,-1.1) node[below] {\footnotesize{$s_{\ast}$}};
    \fill (0.5,-1.1) node[below] {\footnotesize{$\frac{1}{2}$}};
    \fill (1,-1.1) node[below] {\footnotesize{$\overline{\theta}$}};
    \fill (-0.5,0.1) node[above] {\footnotesize{$U(s)$}};
    \fill (-0.5,0) node[left] {\footnotesize{$0$}};
    \fill (-0.5,-1) node[left] {\footnotesize{$-c(1)$}};
    \fill (1,-1.1) node[right] {\footnotesize{$s$}};
    
    \draw[dotted] (-0.5,0)--(1,0);
    \draw[dotted] (-0.5,-1)--(1,-1);
    \draw[dotted] (0.25,-0.5)--(0.25,-1.1);
    \draw[dotted] (0,-1)--(0,-1.1);
    \draw[dotted] (0.5,0)--(0.5,-1.1);
    \draw[line width=0.5pt,blue] (-0.5,-1)--(0,-1)--(0.5,0)--(1,0);
    \draw[line width=2.5pt,red, draw opacity=0.4] (-0.5,-1)--(0,-1)--(0.55,0.1);
	\end{tikzpicture}
\end{subfigure}
\caption{\label{fig:indirect_util_linear}
This figure depicts Proposer's indirect utility function $U(s)$ (blue solid) and the smallest convex function $\phi(s)$ (red translucent) above $U$ when $c$ is linear. The left panel considers the case where $\underline{\theta}\geq 0$, while the other two panels represent the case where $\underline{\theta}<0$. In the middle panel, $\mathbb{E}[\theta|\theta\geq0]>1/2$, while $\mathbb{E}[\theta|\theta\geq0]< 1/2$ in the right panel.}
\end{figure}
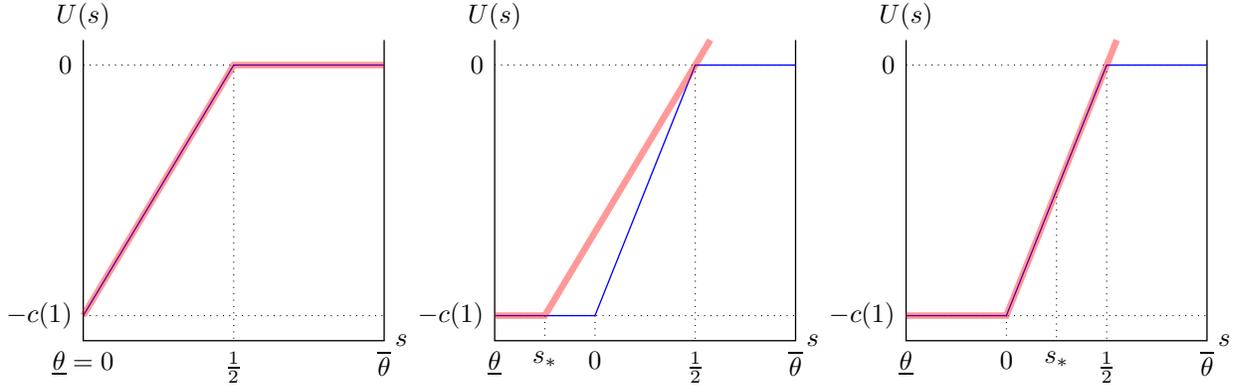

Suppose Proposer's loss function is linear, that is, $c(|1-a|)=|1-a|$. In this case, as depicted in \cref{fig:indirect_util_linear}, $U$ is a piecewise linear function. Clearly, $U$ is concave over $[\underline{\theta},\overline{\theta}]$ if $\underline{\theta}\geq 0$ (the left panel) but not if $\underline{\theta}<0$ (the two right panels). This implies that no information is optimal if and only if $\underline{\theta}\geq 0$. In addition, if $\underline{\theta}<0$ then an optimal signal structure is a binary experiment with cutoff $\min\{0,s_\ast\}$ such that
\begin{align*}
s^{\ast}=\mathbb{E}[\theta|\theta\geq s_{\ast}]=\frac{1}{2}.  
\end{align*}
To interpret the cutoff, note that if $2\mathbb{E}[\theta|\theta \geq 0] <1$ then Proposer is risk neutral on the relevant range and seeks to maximize the expected action, which is achieved by setting $s_*=0$.  Otherwise, an optimal experiment provides just enough information that Vetoer is willing to accept Proposer's ideal policy $1$ upon learning that $\theta\geq s_{\ast}$.

Clearly, conditional on $\underline{\theta}<0$ and $\mathbb{E}[\theta|\theta\geq0]>1/2$, Proposer's optimal (positive) proposal ($2\mathbb{E}[\theta|\theta\geq s_{\ast}]=1$) is independent of the details of $F$. Vetoer accepts $1$ if and only if $\theta\geq s_{\ast}$, whose probability is equal to $1-F(s_{\ast})$. To see how this probability depends on $F$, consider a simple case where $F$ is a uniform distribution over $[\underline{\theta},\overline{\theta}]$ (and $\underline{\theta}<0<\overline{\theta}$). In that case, $\mathbb{E}[\theta|\theta\geq s_{\ast}]=\frac{s_{\ast}+\overline{\theta}}{2}$, and thus
\begin{align*}
    1-F(s_{\ast})=\frac{\overline{\theta}-s_{\ast}}{\overline{\theta}-\underline{\theta}}=\frac{2\overline{\theta}-1}{\overline{\theta}-\underline{\theta}}.
\end{align*}
It is easy to see that this probability is increasing in both $\underline{\theta}$ and $\overline{\theta}$. If $\underline{\theta}<0$ and $\mathbb{E}[\theta|\theta\geq0]<1/2$, then Proposer's optimal (positive) proposal depends on $F$. When $F$ is a uniform distribution, the proposal $2\mathbb{E}[\theta]=\overline{\theta}$ is increasing in $\overline{\theta}$ and independent of $\underline{\theta}$. The probability that Vetoer accepts is $1-F(0)$, which is also increasing in both $\underline{\theta}$ and $\overline{\theta}$.


\subsection{Quadratic Proposer}\label{subsec:quadratic_proposer}

Now suppose Proposer's loss function is quadratic, that is, $c(|1-a|)=(1-a)^{2}$. In this case, \eqref{eq:no_info_opt}---a necessary and sufficient condition for no information to be optimal---reduces to 
\begin{align*}
-1 \leq -(1-2\mathbb{E}[\theta])^2+(4-8\mathbb{E}[\theta])(\underline{\theta}-\mathbb{E}[\theta])\Leftrightarrow
-\underline{\theta}\left(1-2\mathbb{E}[\theta]\right)\leq \mathbb{E}[\theta]^2.
\end{align*}
If $F$ is uniform over $[\underline{\theta},\overline{\theta}]$ then the condition further simplifies to\footnote{Observe that if $\overline{\theta}=1$ then $\kappa(\overline{\theta})=-1/3$, which is consistent with our derivation at the end of \cref{sec:example}.}
\begin{align*}
    \underline{\theta}\left(1-2\frac{\underline{\theta}+\overline{\theta}}{2}\right)\leq \left(\frac{\underline{\theta}+\overline{\theta}}{2}\right)^{2}\Leftrightarrow \underline{\theta}\geq \kappa(\overline{\theta}):=-\frac{\overline{\theta}^{2}}{2-\overline{\theta}+2\sqrt{1-\overline{\theta}+\overline{\theta}^{2}}}.
\end{align*}
That no information is optimal if, and only if, $\underline{\theta}$ is not too negative is clear from \cref{fig:indirect_util_DM} and the discussion following \cref{prop:quadratic_persuade_first}. Through rather tedious calculus, it can be shown that $\kappa$ is strictly decreasing in $\overline{\theta}$. Then, the above condition also suggests that no information is more likely to be optimal the higher $\overline{\theta}$ is. Note that this result does not immediately follow from our characterization in \cref{sec:main}.

Suppose the above condition fails and, therefore, an optimal signal must reveal whether $\theta\geq s_{\ast}$ or not for some $s_{\ast}\in(\underline{\theta},0]$. In this case, \eqref{eq:some_info_opt}---the condition characterizing the optimal $s_{\ast}$---reduces to
\begin{align*}
    \frac{1-(1-2\mathbb{E}[\theta|\theta\geq s_{\ast}])^{2}}{\mathbb{E}[\theta|\theta\geq s_{\ast}]-s_{\ast}}=4(1-2\mathbb{E}[\theta|\theta\geq s_{\ast}]).
\end{align*}
Applying $\mathbb{E}[\theta|\theta\geq s_{\ast}]=\frac{s_{\ast}+\overline{\theta}}{2}$ to this equation, we obtain a quadratic equation for $s_{\ast}$, whose unique negative solution is $s_{\ast}=\kappa(\overline{\theta})$. This implies that if $\underline{\theta}<\kappa(\overline{\theta})$ then it is optimal for Proposer to reveal whether $\theta\geq \kappa(\overline{\theta})$ or not. It follows that Proposer's optimal proposal is $2\mathbb{E}[\theta|\theta\geq s_{\ast}]=\kappa(\overline{\theta})+\overline{\theta}$. This value is clearly independent of $\underline{\theta}$ and can be shown to be increasing in $\overline{\theta}$. It also can be shown that, as in the previous linear case, the probability that Vetoer accepts $\kappa(\overline{\theta})+\overline{\theta}$ increases in both $\underline{\theta}$ and $\overline{\theta}$.

\subsection{General Comparative Statics}\label{subsec:comp_statics}

Comparing the two examples, there is a sense in which Proposer wishes to provide less information in the quadratic case than in the linear case. To begin with, no information is optimal in the linear case only when $\underline{\theta}\geq 0$, but it can be in the quadratic case even when $\underline{\theta}$ is strictly negative (as long as $\underline{\theta}\geq\kappa(\overline{\theta})$). Furthermore, when no information is suboptimal in both cases (i.e., $\underline{\theta}<\kappa(\overline{\theta})$), the relevant cutoff $s_{\ast}$ is always larger in the linear case than in the quadratic case; that is, $1-\overline{\theta}\geq \kappa(\overline{\theta})$. Since a binary signal can be interpreted as to fully reveals $\theta$ if $\theta<s_{\ast}$ but no information if $\theta\geq s_{\ast}$, this also means that Proposer provides less information in the quadratic case. Equivalently, since a positive proposal is made and accepted when $\theta\geq s_{\ast}$, Vetoer is less likely to learn information which causes them to reject the status quo when Proposer has quadratic loss.  Note that maintaining the status quo is also less likely when the bounds on the support shift right. Our general comparative statics result below generalizes these findings and illustrates the driving force. 

We consider changes in $F$ and the concavity of Proposer's utility function (convexity of $c$).  We define more risk averse as a higher co-efficient of absolute risk aversion,\footnote{That is, if $-\frac{u_1^{\prime \prime}(a)}{u_1^{\prime}(a)} \leq -\frac{u_2^{\prime \prime}(a)}{u_2^{\prime}(a)}$ for all $a \in [0, 1]$, then Proposer is more risk averse under $u_2$ than $u_1$.} and show that a more risk averse proposer chooses lower $s_*$ (and $s^*$) to reduce vetoes.   Similarly, if the distribution of $\theta$ shifts right, for any $s_*$ a more favorable proposal, $s^*$, would result. Proposer would then have an incentive to decrease $s_*$ to reduce vetoes, resulting in a higher proposal and fewer vetoes.  The next result establishes this formally.

\begin{proposition}
\label{p:statics}
\begin{enumerate}
\item[(i)] $s_*$ and $s^*$ are decreasing in Proposer's risk aversion.
\item[(ii)] A rightward likelihood ratio shift in the distribution of $\theta$ decreases $s_*$ and $F(s_*)$ and increases $s^*$.
\end{enumerate}
\end{proposition}

Note that likelihood ratio dominance is stronger than needed to get the second result. In fact, the result holds as long as $\mathbb{E}[\theta|\theta \geq s]$ becomes larger for any $s$; this is weaker than likelihood ratio dominance but not implied by first order stochastic dominance.

\section{Vetoer Linear Loss}\label{sec:linear_loss}

So far, we have assumed that Vetoer has a quadratic loss function (i.e., his ex post payoff is $-(\theta-a)^{2}$ when his ideal point is $\theta$ and the chosen policy is $a$). This section studies the extent to which our results depend on this specification. In particular, we characterize Proposer's optimal policies when Vetoer has a linear loss function (i.e., his ex post payoff is given by $-|\theta-a|$). 

\subsection{Vetoer's Acceptance Decision}\label{subsec:linear_vetoer_accept}

As in \cref{sec:vetoer_accept}, suppose Vetoer's posterior belief about $\theta$ is represented by the distribution $G$. When his ex post payoff is $-|\theta-a|$, he is willing to accept proposal $p$ if and only if 
\begin{align*}
    V(p,G)=-\int|\theta-p|dG(\theta)\geq V(0,G)=-\int|\theta|dG(\theta),
\end{align*}
which can be rewritten as 
\begin{align*}
    \int_{\underline{\theta}}^{p}(p-\theta)dG(\theta)+\int_{p}^{\overline{\theta}}(\theta-p)dG(\theta)\leq \int_{\underline{\theta}}^{\max\{\underline{\theta},0\}}(-\theta)dG(\theta)+\int_{\max\{\underline{\theta},0\}}^{\overline{\theta}}\theta dG(\theta).
\end{align*}
Unlike in our baseline quadratic case, this condition cannot be reduced to a more tractable and easy-to-interpret condition. 


To proceed, we focus on the case where $\theta$ is either $\ell$ or $h$, and $0\leq\ell<h$.\footnote{This latter assumption is not necessary for the subsequent analysis; with a linear loss, $\ell<0$ and $\ell=0$ are equivalent.  However it reduces the number of cases to consider and, more importantly, highlights differences from our baseline quadratic case.}  Letting $\mu$ denote Vetoer's belief that his ideal point is $h$, the above acceptance condition becomes
\begin{equation*}
    V(p,\mu)=-(1-\mu)|p-\ell|-\mu |p-h|\geq  V(0,\mu)=-(1-\mu)|\ell|-\mu |h|,
\end{equation*}
which can be simplified to 
\begin{equation*}
    \mu\geq \phi(p):=\frac{p-2\ell}{2\left(\min\{p,h\}-\ell\right)}.
\end{equation*}
If $p$ is around $2\ell$ then $\phi(p)$ is close to $0$, so this inequality necessarily holds. In contrast, it fails if $p$ is sufficiently large. 

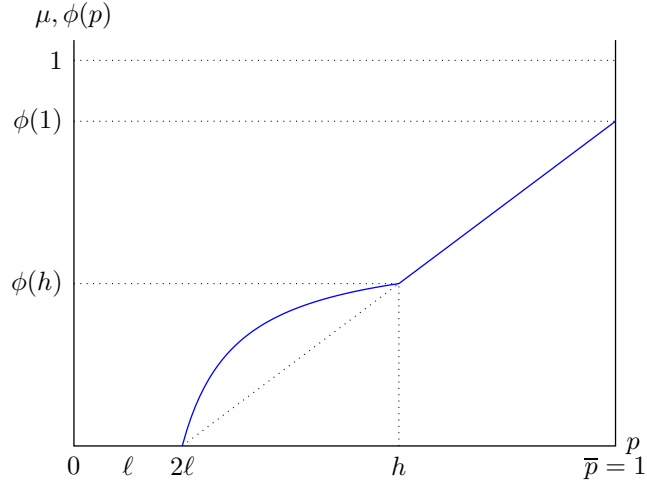
\begin{figure}
\begin{center}
\begin{tikzpicture}[scale=0.9]
		
    \draw[line width=0.5pt] (0,6)--(0,0)--(8,0)--(8,6); 
	
	\fill (8,0) node[right] {\footnotesize{$p$}};
    \fill (8,0) node[below] {\footnotesize{$\overline{p}=1$}};
    \fill (0,0) node[below] {\footnotesize{$0$}};
	\fill (0,6) node[above] {\footnotesize{$\mu,\phi(p)$}};

    \fill (0,5.7) node[left] {\footnotesize{$1$}};
    
    \draw[line width=0.5pt,blue] plot[smooth]
    (1.6,0)--(1.6653,0.22642)--(1.7306,0.42105)--(1.7959,0.59016)--(1.8612,0.73846)--(1.9265,0.86957)--(1.9918,0.9863)--(2.0571,1.0909)--(2.1224,1.1852)--(2.1878,1.2706)--(2.2531,1.3483)--(2.3184,1.4194)--(2.3837,1.4845)--(2.449,1.5446)--(2.5143,1.6)--(2.5796,1.6514)--(2.6449,1.6991)--(2.7102,1.7436)--(2.7755,1.7851)--(2.8408,1.824)--(2.9061,1.8605)--(2.9714,1.8947)--(3.0367,1.927)--(3.102,1.9574)--(3.1673,1.9862)--(3.2327,2.0134)--(3.298,2.0392)--(3.3633,2.0637)--(3.4286,2.087)--(3.4939,2.1091)--(3.5592,2.1302)--(3.6245,2.1503)--(3.6898,2.1695)--(3.7551,2.1878)--(3.8204,2.2054)--(3.8857,2.2222)--(3.951,2.2383)--(4.0163,2.2538)--(4.0816,2.2687)--(4.1469,2.2829)--(4.2122,2.2967)--(4.2776,2.3099)--(4.3429,2.3226)--(4.4082,2.3348)--(4.4735,2.3467)--(4.5388,2.3581)--(4.6041,2.3691)--(4.6694,2.3797)--(4.7347,2.39)--(4.8,2.4);

    \draw[line width=0.5pt,blue] (4.8,2.4)--(8,4.8); 

    \fill (0.8,0) node[below] {\footnotesize{$\ell$}};
    \fill (1.6,0) node[below] {\footnotesize{$2\ell$}};
    \fill (4.8,0) node[below] {\footnotesize{$h$}};
    
    \draw[dotted] (1.6,0)--(4.8,2.4);
    \draw[dotted] (0,5.7)--(8,5.7); 
    \draw[dotted] (0,4.8)--(8,4.8);
    \draw[dotted] (0,2.4)--(4.8,2.4);
    \draw[dotted] (4.8,0)--(4.8,2.4); 
    
    \fill (0,4.8) node[left] {\footnotesize{$\phi(1)$}};
    \fill (0,2.4) node[left] {\footnotesize{$\phi(h)$}};





    


    

	\end{tikzpicture}
\end{center}\caption{\label{fig:phi_examples}This figure shows $\phi(p)$---the lowest belief level at which Vetoer is willing to accept $p$---in the binary linear vetoer case.} 
\end{figure}

\cref{fig:phi_examples} depicts the representative shape of the function $\phi$. Particularly notable is that $\phi$ has an upward kink at $h$, that is, $\lim_{p\to h-}\phi^{\prime}(p)<\lim_{p\to h+}\phi^{\prime}(p)$.\footnote{Note that this is relevant only when $h>2\ell$; otherwise, $\phi$ is a linear function above $2\ell$.} It stems from the single-peakedness of Vetoer's preferences. To see this more clearly, consider a marginal increase of Proposer's proposal $p\in[2\ell,1)$. If $p\geq h$ then the increase is detrimental to Vetoer, regardless of his ideal point $\theta$. If $p<h$, however, it is beneficial to Vetoer when his ideal point is $h$, although it is still harmful when his ideal point is $\ell$. This implies that as $p$ rises, a relatively smaller increase of $\mu$ is sufficient for Vetoer to stay indifferent between accepting and rejecting $p$ when $p<h$ than when $p>h$. The kink at $h$ reflects this change around $h$. In fact, the same effect is present in our baseline quadratic case. However, in that case, Vetoer's ex post payoff is smooth around his bliss point, so it does not result in a kink at $h$. 

It is convenient to define the inverse of the function $\phi$. Formally, for each $\mu\in[0,1]$, we let 
\begin{equation*}
    \psi(\mu):=\max\{p\in[0,1]:\phi(p)\leq\mu\}. 
\end{equation*}
Since $\phi$ is increasing, $\psi(\mu)$ represents the highest proposal below $1$ that Vetoer would accept with belief $\mu$. In other words, Vetoer accepts $p$ with belief $\mu$ if and only if $p\leq\psi(\mu)$.


\subsection{Persuasion-First Model}\label{sec:linear_persuade_first} 

As in \cref{sec:main}, we first consider the persuasion-first model. As illustrated above, given his belief $\mu$, Vetoer accepts $p$ if and only if $p\leq\psi(\mu)$. Then, clearly, Proposer's optimal proposal is $\psi(\mu)$, and Proposer's indirect utility associated with posterior $\mu$ is given by 
\begin{equation*}
    \widehat{U}(\mu)=-c\left(1-\psi(\mu)\right).
\end{equation*}

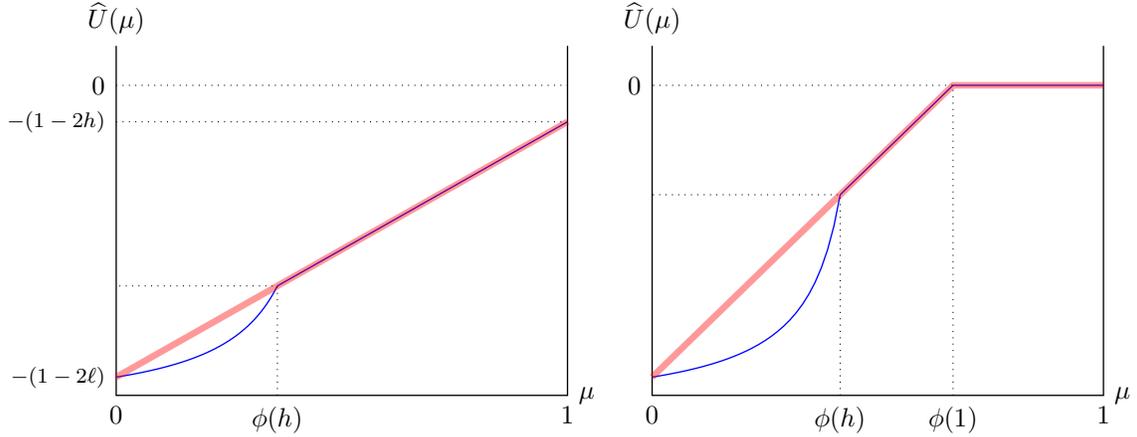
\begin{figure}
\begin{center}
\begin{tikzpicture}[scale=0.75]
		
    \draw[line width=0.5pt] (0,6.2)--(0,0)--(8,0)--(8,6.2);
	
	\fill (8,0) node[right] {\footnotesize{$\mu$}};
    \fill (8,0) node[below] {\footnotesize{$1$}};
    \fill (0,0) node[below] {\footnotesize{$0$}};
	\fill (0,6.2) node[above] {\footnotesize{$\widehat{U}(\mu)$}};


    \fill (0,0.32353) node[left] {\scriptsize{$-(1-2\ell)$}};

    \draw[line width=0.5pt,blue] plot[smooth]
    (0,0.32353)--(0.057143,0.33291)--(0.11429,0.34256)--(0.17143,0.3525)--(0.22857,0.36275)--(0.28571,0.3733)--(0.34286,0.38419)--(0.4,0.39542)--(0.45714,0.40702)--(0.51429,0.419)--(0.57143,0.43137)--(0.62857,0.44417)--(0.68571,0.4574)--(0.74286,0.4711)--(0.8,0.48529)--(0.85714,0.5)--(0.91429,0.51525)--(0.97143,0.53108)--(1.0286,0.54751)--(1.0857,0.56459)--(1.1429,0.58235)--(1.2,0.60084)--(1.2571,0.6201)--(1.3143,0.64018)--(1.3714,0.66113)--(1.4286,0.68301)--(1.4857,0.70588)--(1.5429,0.72982)--(1.6,0.7549)--(1.6571,0.78121)--(1.7143,0.80882)--(1.7714,0.83786)--(1.8286,0.86842)--(1.8857,0.90064)--(1.9429,0.93464)--(2,0.97059)--(2.0571,1.0087)--(2.1143,1.049)--(2.1714,1.0919)--(2.2286,1.1376)--(2.2857,1.1863)--(2.3429,1.2383)--(2.4,1.2941)--(2.4571,1.354)--(2.5143,1.4186)--(2.5714,1.4882)--(2.6286,1.5637)--(2.6857,1.6458)--(2.7429,1.7353)--(2.8,1.8333)--(2.8571,1.9412);

    \draw[line width=0.5pt,blue] (2.8571,1.9412)--(8,4.8529);

    \draw[dotted] (0,5.5)--(8,5.5);

    \draw[dotted] (0,4.8529)--(8,4.8529);

    \fill (0,4.8529) node[left] {\scriptsize{$-(1-2h)$}};

    \fill (0,5.5) node[left] {\footnotesize{$0$}};

    \fill (2.8571,0) node[below] {\footnotesize{$\phi(h)$}};

    \draw[dotted] (2.8571,0)--(2.8571,1.9412)--(0,1.9412);

    \draw[line width=2.5pt,red,draw opacity=0.4] (0,0.32353)--(8,4.8529);

    \begin{scope}[xshift=9.5cm]

    \draw[line width=0.5pt] (0,6.2)--(0,0)--(8,0)--(8,6.2);
	
	\fill (8,0) node[right] {\footnotesize{$\mu$}};
    \fill (8,0) node[below] {\footnotesize{$1$}};
    \fill (0,0) node[below] {\footnotesize{$0$}};
	\fill (0,6.2) node[above] {\footnotesize{$\widehat{U}(\mu)$}};


    \draw[line width=0.5pt,blue] plot[smooth]
    (0,0.32353)--(0.11111,0.34202)--(0.22222,0.36159)--(0.33333,0.38235)--(0.44444,0.40441)--(0.55556,0.42789)--(0.66667,0.45294)--(0.77778,0.47972)--(0.88889,0.5084)--(1,0.53922)--(1.1111,0.5724)--(1.2222,0.60824)--(1.3333,0.64706)--(1.4444,0.68926)--(1.5556,0.73529)--(1.6667,0.78571)--(1.7778,0.84118)--(1.8889,0.90248)--(2,0.97059)--(2.1111,1.0467)--(2.2222,1.1324)--(2.3333,1.2294)--(2.4444,1.3403)--(2.5556,1.4683)--(2.6667,1.6176)--(2.7778,1.7941)--(2.8889,2.0059)--(3,2.2647)--(3.1111,2.5882)--(3.2222,3.0042)--(3.3333,3.5588);

    \draw[line width=0.5pt,blue] (3.3333,3.5588)--(5.3333,5.5)--(8,5.5);

    \fill (0,5.5) node[left] {\footnotesize{$0$}};

    \fill (3.3333,0) node[below] {\footnotesize{$\phi(h)$}};

    \fill (5.3333,0) node[below] {\footnotesize{$\phi(1)$}};

    \draw[dotted] (3.3333,0)--(3.3333,3.5588)--(0,3.5588);
    \draw[dotted] (5.3333,0)--(5.3333,5.5)--(0,5.5);



    \draw[line width=2.5pt,red,draw opacity=0.4] (0,0.32353)--(5.3333,5.5)--(8,5.5);
    \end{scope}

	\end{tikzpicture}
\end{center}\caption{\label{fig:Proposer_indirect_linear}Both panels show Proposer's indirect utility function $\widehat{U}(\mu)$ (blue solid) and its smallest concave closure $\overline{U}(\mu)$ (red translucent) when $c$ is linear. In both panels, $\ell=0.1$. In the left panel $h=0.45(<1/2)$, while in the right panel $h=0.7(>1/2)$.}
\end{figure}

If Proposer chooses a distribution $H$ of posteriors then her expected payoff is given by $\mathbb{E}[\widehat{U}(\mu)]=\int \widehat{U}(\mu)dH(\mu)$. This implies that Proposer's optimal persuasion problem can be written as
\begin{equation*}
    \max_{H\in\Delta([0,1])}\int \widehat{U}(\mu)dH(\mu)\text{ subject to }\int \mu dH(\mu)=\mu_{0},
\end{equation*}
where $\Delta([0,1])$ denotes the set of all distributions over $[0,1]$. As observed by \citet{KG2011}, this linear programming problem can be analyzed by the method of concavification \citep{AM1995}. Specifically, the optimal value, denoted by $\overline{U}(\mu_{0})$, coincides with the value of the smallest concave function uniformly above $\widehat{U}$; see the red translucent curves in \cref{fig:Proposer_indirect_linear}. Any distribution $H$ such that $supp(H)\subseteq\{\mu:\widehat{U}(\mu)=\overline{U}(\mu)\}$, $\int \mu dH(\mu)=\mu_{0}$, and $\int \widehat{U}(\mu)dH(\mu)=\overline{U}(\mu_{0})$ is an optimal strategy. 

\paragraph{No information.} We begin by studying the condition under which Proposer does not wish to provide any information. 


\begin{proposition}\label{prop:no_info_optimal}
Consider the persuasion-first model in which Vetoer's loss function is linear and his bliss point is either $\ell$ or $h$. Then no information is optimal if $\mu_{0}\geq\max\{0,\phi(h)\}$. 
Conversely, there exists $\overline{\ell}(>0)$ such that for any $\ell<\overline{\ell}$, no information is sub-optimal if $\mu_{0}$ is sufficiently small. 
\end{proposition}

To understand this result, recall that $\psi$ is convex below $\phi(h)$ and piece-wise linear and concave above $\phi(h)$; see the blue curves in \cref{fig:Proposer_indirect_linear}. Combining this with the fact that $\widehat{U}(\mu)=-c(1-\psi(\mu))$, it follows that, although $\widehat{U}$ may not be convex or concave below $\phi(h)$, it is unambiguously concave above $\phi(h)$. The result is straightforward from this shape of $\widehat{U}$. Intuitively, above $\phi(h)$, there is no preference alignment between Proposer and Vetoer; an increase of $p$ benefits the former but hurts the latter. Then, Vetoer's linear (risk-neutral) preferences imply that Proposer cannot increase the expected value of the chosen policy. She then prefers the ``safest'' strategy inducing a deterministic outcome to any other non-degenerate strategy. 

The second result follows from the fact that if $\ell$ is close to $0$ then $\widehat{U}(\mu)=-c(1-\psi(\mu))$ is convex in the small neighborhood of $\mu$, regardless of convexity of $c$. To see this more clearly, observe that 
\begin{equation*}
    \widehat{U}^{\prime}(\mu)=c^{\prime}(1-\psi(\mu))\psi^{\prime}(\mu)>0.
\end{equation*}
If $\ell$ is close to $0$ then $\widehat{U}^{\prime}(0)=c^{\prime}(1-2\ell)2\ell\approx 0$. It then follows that that $\widehat{U}^{\prime}(\mu)$ must be increasing---that is, $\widehat{U}(\mu)$ is convex---around $0$. 

\paragraph{Optimal persuasion.} In general, a necessary and sufficient condition for no information to be optimal (equivalently, for persuasion to have value) and the structure of an optimal signal depend on the detailed shape of $c$. We illustrate this with the simplest example in which Proposer's loss function is also linear. The analysis can be easily extended for other loss functions. 


\begin{example}[Linear loss]\label{ex:persuade_first_linear}
    If $c$ is linear (i.e., $c(|1-a|)=|1-a|)$ then
    \begin{equation*}
        \widehat{U}(\mu)=-1+\psi(\mu)=-1+\left\{
    \begin{array}{ll}
    2\ell\frac{(1-\mu)}{1-2\mu}&\text{if }\mu\leq\phi(h)\\
    2\left((1-\mu)\ell+\mu h\right)&\text{if }\mu\in\left[\phi(h),\min\{1,\phi(1)\}\right]\\
    1&\text{if }p\in[\phi(1),1].
    \end{array}
    \right.
    \end{equation*}
    As depicted in \cref{fig:Proposer_indirect_linear}, $\widehat{U}$ is strictly convex below $\phi(h)$ and so coincides with $\overline{U}$ only above $\phi(h)$. This implies that $\mu\geq\phi(h)$ is necessary and sufficient for no information to be optimal. In addition, if $\mu<\phi(h)$ then any signal such that the support of posteriors belongs to $0\cup[\phi(h),\min\{\phi(1),1\}]$ is optimal. For example, the binary signal that induces posterior $0$ or $\phi(h)$ is optimal.
\end{example}

Note that the optimal persuasion mechanism can be achieved with a binary message as in the quadratic model.  However this emerges only because $\theta$ can take on only two values.  As illustrated in \cref{three}, more than two messages may be necessary to achieve the optimal when Vetoer has linear loss.  

\subsection{Proposal-First Model}\label{subsec:linear_propose_first}

In our baseline quadratic case, the persuasion-first optimal outcome could always be implemented in the proposal-first model. The result no longer holds in the current linear environment. Hence, we provide a separate analysis and compare the resulting outcome to the persuasion-first one. 

\begin{figure}
\begin{center}
\begin{tikzpicture}[scale=1]
		
	\draw[line width=0.5pt] (0,4.5)--(0,0)--(6.5,0)--(6.5,4.5); 
	
	\fill (6.5,0) node[right] {\footnotesize{$\mu$}};
    \fill (6.5,0) node[below] {\footnotesize{$1$}};
    \fill (0,0) node[below] {\footnotesize{$0$}};
	\fill (0,4.5) node[above] {\footnotesize{$\widetilde{U}(\mu;p)$}};

    \draw[dotted] (0,3.5)--(6.5,3.5); 
    \draw[dotted] (3,0)--(3,3.5); 
    \draw[dotted] (3,0.5)--(6.5,0.5); 
    
    \draw[line width=1pt] (0,0.5)--(3,0.5);
    \draw[line width=1pt,dashed] (3,0.5)--(3,3.5);
    \draw[line width=1pt] (3,3.5)--(6.5,3.5);
    
    \fill (3,0) node[below] {\footnotesize{$\phi(p)$}};
    \fill (0,3.5) node[left] {\footnotesize{$-c(1-p)$}};
    \fill (0,0.5) node[left] {\footnotesize{$-c(1)$}};

    \draw[line width=2.5pt,red, draw opacity=0.4] (0,0.5)--(3,3.5)--(6.5,3.5); 

    \fill (1.5,0) node[below] {\footnotesize{$\mu_{0}$}};

    \draw[fill=blue] (0,0.5) circle (2pt);
    \draw[fill=blue] (3,3.5) circle (2pt);

    \draw[dotted] (1.5,0)--(1.5,2); 
    \draw[fill=red] (1.5,2) circle (2pt);

    
	\end{tikzpicture}
\end{center}\caption{\label{fig:concavification}This figure illustrates the optimal information structure given $p$. In the figure, the black solid curve denotes Proposer's indirect utility as a function of posterior $\mu$, while the red translucent curve represents the smallest concave closure of the indirect utility function.} 
\end{figure}
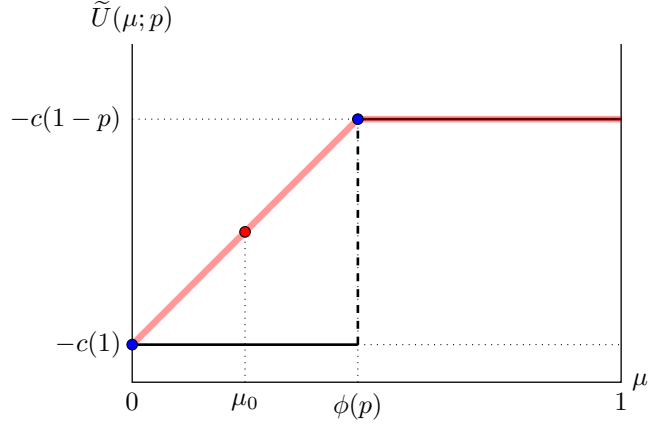

Suppose Proposer has chosen $p\in[\underline{p},\overline{p}]$. To solve for her optimal persuasion strategy, recall that Vetoer accepts $p$ if and only if $\mu\geq\phi(p)$. This implies that, as depicted in \cref{fig:concavification}, Proposer's utility as a function of the induced posterior $\mu$ is a step function, with a jump at $\phi(p)$ (the black solid curve). In this case, the following is immediate from the standard concavification argument:  
\begin{itemize}
    \item If $\mu_{0}\geq \phi(p)$ then it is optimal for Proposer to provide no information. 
    \item If $\mu_{0}<\phi(p)$ then a binary signal that leads to posterior $0$ or $\phi(p)$ is optimal. 
\end{itemize}

The best-case scenario for Proposer is that Vetoer accepts $\overline{p}:=\min\{2h,1\}$ with probability $1$. The above result implies that Proposer obtains this best outcome if and only if 
\begin{equation}\label{eq:mu0_large}
\mu_{0}\geq\phi(\overline{p})=\frac{\overline{p}-2\ell}{2(h-\ell)}.
\end{equation}
Notice that $\phi(\overline{p})<1$ if $h>1/2$, but $\phi(\overline{p})=1$ if $h\leq 1/2$. This suggests that if $h>1/2$ then Proposer can implement $\overline{p}$ with probability $1$ for $\mu_{0}$ sufficiently large; while if $h\leq 1/2$ then $\overline{p}=2h$ cannot be implemented with probability $1$. Intuitively, Vetoer is indifferent between $0$ and $\overline{p}=2h(\leq 1)$ if he is certain that his ideal point is $h$. If he assigns a positive probability to $\ell$ then he prefers $0$ to $2h$. From now on, we focus on the case where \eqref{eq:mu0_large} fails.

In general, Proposer's (indirect) expected utility as a function of proposal $p$ is given by 
\begin{equation*}
\widetilde{U}(p)=:-\min\left\{\frac{\mu_{0}}{\phi(p)},1\right\}c(1-p)-\max\left\{\frac{\phi(p)-\mu_{0}}{\phi(p)},0\right\}c(1).
\end{equation*}
Since $\phi(p)\leq\mu_{0}\Leftrightarrow p\leq\psi(\mu_{0})$, $\widetilde{U}$ can be rewritten as 
\begin{equation*}
\widetilde{U}(p)=
\left\{\begin{array}{ll}
-c(1-p)&\text{if }p\in[\underline{p},\psi(\mu_{0}))\\
-c(1)+\frac{\mu_{0}}{\phi(p)}(c(1)-c(1-p))&\text{if }p\in[\psi(\mu_{0}),\overline{p}].
\end{array}\right.
\end{equation*}

\paragraph{Optimal proposal.} It suffices to find $p$ that maximizes $\widetilde{U}(p)$. Clearly, $\widetilde{U}$ is increasing below $\psi(\mu_{0})$. In addition, it is decreasing after $\max\{\psi(\mu_{0}),h\}$.\footnote{To see this, observe that for $p>\max\{\psi(\mu_{0}),h\}$, 
\begin{eqnarray*}
    \widetilde{U}^{\prime}(p)&=&\frac{\mu_{0}}{\phi(p)^{2}}\left[c^{\prime}(1-p)\phi(p)-(c(1)-c(1-p))\phi^{\prime}(p)\right]=\frac{\mu_{0}}{\phi(p)^{2}2(h-\ell)}\left[c^{\prime}(1-p)(p-2\ell)-(c(1)-c(1-p))\right]
\end{eqnarray*}
This expression is strictly negative because $\ell\geq 0$ and $c$ is convex, so 
\begin{equation*}
    c^{\prime}(1-p)(p-2\ell)<c^{\prime}(1-p)p=c^{\prime}(1-p)\left(1-(1-p)\right)\leq c(1)-c(1-p).
\end{equation*}} For $p\in[\psi(\mu_{0}),h]$, $\widetilde{U}(p)$ is given by 
\begin{equation*}
    \widetilde{U}(p)=-c(1)+\mu_{0}(c(1)-c(1-p))\left(1+\frac{\ell}{p-2\ell}\right). 
\end{equation*}
This function may not be monotone, as $c(1)-c(1-p)$ is strictly increasing in $p$, while $\frac{\ell}{p-\ell}$ is strictly decreasing in $p$. We make a mild technical assumption on $c$ that $\widetilde{U}(p)$ is quasi-convex when extended over $(2\ell,\infty)$; this is plausible, because the value of this function grows arbitrarily large as $p$ tends to $2\ell$ or $\infty$, but its precise shape depends on the detail of $c$. It is easy to see that this assumption always holds if $c$ is linear or quadratic. 

\begin{proposition}\label{prop:propose_first_optimal}
Consider the proposal-first model in which Vetoer's loss function is linear and Vetoer's bliss point is either $\ell$ or $h$. Suppose $(c(1)-c(1-p))(p-\ell)/(p-2\ell)$ is quasi-convex in $p(>\underline{p})$ and \eqref{eq:mu0_large} fails. 
\begin{itemize}
    \item Proposing $h$, with a binary signal that leads to posterior $0$ or $\phi(h)$, is optimal if 
        \begin{equation}\label{eq:h_optimal_cond}
            -c(1-\psi(\mu_{0}))\leq -c(1)+\frac{\mu_{0}}{\phi(h)}(c(1)-c(1-h)).\footnote{For the case of linear loss, \eqref{eq:h_optimal_cond} simplifies to
\begin{equation*}
    \frac{\ell}{h}\leq\mu_{0}\leq\phi(h)=\frac{h-2\ell}{2(h-\ell)}.
\end{equation*}}
        \end{equation}
    \item Otherwise, proposing $\psi(\mu_{0})$, with no information, is optimal. 
\end{itemize}
\end{proposition}

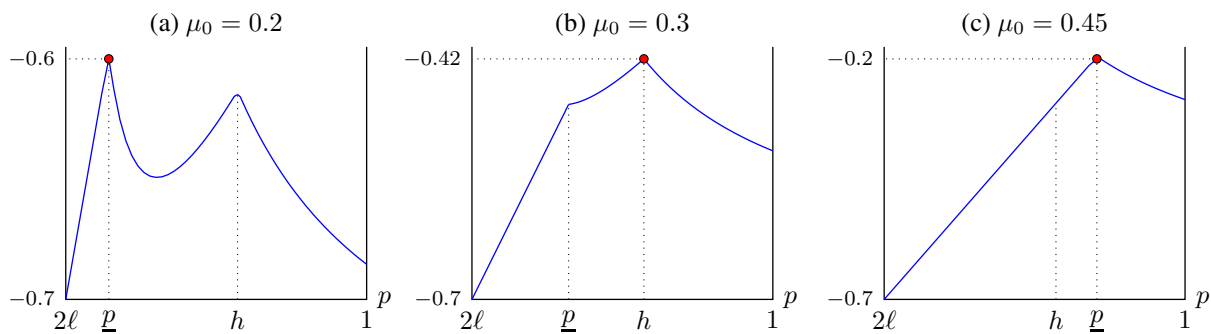
\begin{figure}
\begin{center}
\begin{tikzpicture}[scale=0.8]
		
    \draw[line width=0.5pt] (0,4.2)--(0,0)--(5,0)--(5,4.2); 
     
    \fill (5,0) node[right] {\footnotesize{$p$}};
    \fill (5,0) node[below] {\footnotesize{$1$}};
    \fill (0,0) node[below] {\footnotesize{$2\ell$}};

    \draw[line width=0.5pt,blue] plot[smooth]
    (0,0)--(0.1,0.57143)--(0.2,1.1429)--(0.3,1.7143)--(0.4,2.2857)--(0.5,2.8571)--(0.6,3.4286)--(0.7157,4); 

    \draw[line width=0.5pt,blue] plot[smooth]
    (0.7157,4)--(0.8,3.4904)--(0.9,2.9901)--(1,2.6356)--(1.1,2.3871)--(1.2,2.2181)--(1.3,2.1102)--(1.4,2.0505)--(1.5,2.0292)--(1.6,2.0391)--(1.7,2.0747)--(1.8,2.1318)--(1.9,2.2069)--(2,2.2974)--(2.1,2.401)--(2.2,2.516)--(2.3,2.6408)--(2.4,2.7743)--(2.5,2.9155)--(2.6,3.0633)--(2.7,3.2171)--(2.8,3.3763)--(2.8550,3.4079); 

    \draw[line width=0.5pt,blue] plot[smooth]
    (2.8550,3.4079)--(2.9,3.3699)--(3,3.1487)--(3.1,2.9418)--(3.2,2.7478)--(3.3,2.5656)--(3.4,2.3941)--(3.5,2.2324)--(3.6,2.0797)--(3.7,1.9352)--(3.8,1.7984)--(3.9,1.6685)--(4,1.5452)--(4.1,1.4279)--(4.2,1.3161)--(4.3,1.2096)--(4.4,1.1079)--(4.5,1.0107)--(4.6,0.91773)--(4.7,0.82873)--(4.8,0.74344)--(4.9,0.66163)--(5,0.58309);

    \draw[dotted] (0.7157,0)--(0.7157,4)--(0,4);
    \fill (0.7157,0) node[below] {\footnotesize{$\underline{p}$}};
    
    \draw[dotted] (2.8550,0)--(2.8550,3.4079); 
    \fill (2.8550,0) node[below] {\footnotesize{$h$}};

    \fill (0,0) node[left] {\scriptsize{$-0.7$}};

    \fill (0,4) node[left] {\scriptsize{$-0.6$}};

    \fill (2.5,4.2) node[above] {\footnotesize{(a) $\mu_{0}=0.2$}};

    \draw[fill=red] (0.7157,4) circle (2pt);
    
    \begin{scope}[xshift=6.75cm]
    \draw[line width=0.5pt] (0,4.2)--(0,0)--(5,0)--(5,4.2); 
     
    \fill (5,0) node[right] {\footnotesize{$p$}};
    \fill (5,0) node[below] {\footnotesize{$1$}};
    \fill (0,0) node[below] {\footnotesize{$2\ell$}};

    \draw[line width=0.5pt,blue] plot[smooth]
    (0,0)--(0.042017,0.084791)--(0.084034,0.16958)--(0.12605,0.25437)--(0.16807,0.33916)--(0.21008,0.42395)--(0.2521,0.50874)--(0.29412,0.59353)--(0.33613,0.67833)--(0.37815,0.76312)--(0.42017,0.84791)--(0.46218,0.9327)--(0.5042,1.0175)--(0.54622,1.1023)--(0.58824,1.1871)--(0.63025,1.2719)--(0.67227,1.3567)--(0.71429,1.4414)--(0.7563,1.5262)--(0.79832,1.611)--(0.84034,1.6958)--(0.88235,1.7806)--(0.92437,1.8654)--(0.96639,1.9502)--(1.0084,2.035)--(1.0504,2.1198)--(1.0924,2.2046)--(1.1345,2.2893)--(1.1765,2.3741)--(1.2185,2.4589)--(1.2605,2.5437)--(1.3025,2.6285)--(1.3445,2.7133)--(1.3866,2.7981)--(1.4286,2.8829)--(1.4706,2.9677)--(1.5126,3.0525)--(1.5546,3.1373)--(1.5966,3.222)--(1.6066,3.2429);

    \draw[line width=0.5pt,blue] plot[smooth]
    (1.6066,3.2429)--(1.6387,3.2481)--(1.6807,3.2566)--(1.7227,3.2671)--(1.7647,3.2796)--(1.8067,3.2938)--(1.8487,3.3097)--(1.8908,3.3272)--(1.9328,3.3461)--(1.9748,3.3664)--(2.0168,3.3879)--(2.0588,3.4107)--(2.1008,3.4345)--(2.1429,3.4595)--(2.1849,3.4854)--(2.2269,3.5123)--(2.2689,3.54)--(2.3109,3.5686)--(2.3529,3.598)--(2.395,3.6281)--(2.437,3.659)--(2.479,3.6905)--(2.521,3.7227)--(2.563,3.7555)--(2.605,3.7889)--(2.6471,3.8228)--(2.6891,3.8573)--(2.7311,3.8923)--(2.7731,3.9277)--(2.8151,3.9636)--(2.8571,4);

    \draw[line width=0.5pt,blue] plot[smooth]
    (2.8571,4)--(2.8992,3.9483)--(2.9412,3.8981)--(2.9832,3.8493)--(3.0252,3.8018)--(3.0672,3.7556)--(3.1092,3.7107)--(3.1513,3.667)--(3.1933,3.6245)--(3.2353,3.583)--(3.2773,3.5426)--(3.3193,3.5033)--(3.3613,3.4649)--(3.4034,3.4274)--(3.4454,3.3909)--(3.4874,3.3553)--(3.5294,3.3205)--(3.5714,3.2865)--(3.6134,3.2533)--(3.6555,3.2209)--(3.6975,3.1892)--(3.7395,3.1582)--(3.7815,3.1279)--(3.8235,3.0983)--(3.8655,3.0693)--(3.9076,3.041)--(3.9496,3.0132)--(3.9916,2.9861)--(4.0336,2.9595)--(4.0756,2.9334)--(4.1176,2.9079)--(4.1597,2.8829)--(4.2017,2.8584)--(4.2437,2.8344)--(4.2857,2.8108)--(4.3277,2.7877)--(4.3697,2.7651)--(4.4118,2.7429)--(4.4538,2.7211)--(4.4958,2.6997)--(4.5378,2.6787)--(4.5798,2.6581)--(4.6218,2.6378)--(4.6639,2.618)--(4.7059,2.5985)--(4.7479,2.5793)--(4.7899,2.5605)--(4.8319,2.542)--(4.8739,2.5238)--(4.916,2.5059)--(4.958,2.4883)--(5,2.471);

    \draw[dotted] (1.6066,0)--(1.6066,3.2429);
    \fill (1.6066,0) node[below] {\footnotesize{$\underline{p}$}};
    
    \draw[dotted] (2.8571,0)--(2.8571,4)--(0,4); 
    \fill (2.8571,0) node[below] {\footnotesize{$h$}};

    \fill (0,0) node[left] {\scriptsize{$-0.7$}};

    \fill (0,4) node[left] {\scriptsize{$-0.42$}};

    \fill (2.5,4.2) node[above] {\footnotesize{(b) $\mu_{0}=0.3$}};

    \draw[fill=red] (2.8571,4) circle (2pt);
    
    \end{scope}

    \begin{scope}[xshift=13.6cm]
    \draw[line width=0.5pt] (0,4.2)--(0,0)--(5,0)--(5,4.2); 
     
    \fill (5,0) node[right] {\footnotesize{$p$}};
    \fill (5,0) node[below] {\footnotesize{$1$}};
    \fill (0,0) node[below] {\footnotesize{$2\ell$}};

    \draw[line width=0.5pt,blue] plot[smooth]
    (0,0)--(0.1,0.11429)--(0.2,0.22857)--(0.3,0.34286)--(0.4,0.45714)--(0.5,0.57143)--(0.6,0.68571)--(0.7,0.8)--(0.8,0.91429)--(0.9,1.0286)--(1,1.1429)--(1.1,1.2571)--(1.2,1.3714)--(1.3,1.4857)--(1.4,1.6)--(1.5,1.7143)--(1.6,1.8286)--(1.7,1.9429)--(1.8,2.0571)--(1.9,2.1714)--(2,2.2857)--(2.1,2.4)--(2.2,2.5143)--(2.3,2.6286)--(2.4,2.7429)--(2.5,2.8571)--(2.6,2.9714)--(2.7,3.0857)--(2.8,3.2)--(2.9,3.3143)--(3,3.4286)--(3.1,3.5429)--(3.2,3.6571)--(3.3,3.7714)--(3.4,3.8857)--(3.5350,4);
    
    \draw[line width=0.5pt,blue] plot[smooth]
    (3.5350,4)--(3.6,3.9971)--(3.7,3.9321)--(3.8,3.8705)--(3.9,3.8121)--(4,3.7566)--(4.1,3.7038)--(4.2,3.6535)--(4.3,3.6055)--(4.4,3.5598)--(4.5,3.516)--(4.6,3.4742)--(4.7,3.4342)--(4.8,3.3958)--(4.9,3.359)--(5,3.3236);

    \draw[dotted] (3.5350,0)--(3.5350,4)--(0,4);
    \fill (3.535,0) node[below] {\footnotesize{$\underline{p}$}};
    
    \draw[dotted] (2.8550,0)--(2.8550,3.2306); 
    \fill (2.8550,0) node[below] {\footnotesize{$h$}};

    \fill (0,0) node[left] {\scriptsize{$-0.7$}};

    \fill (0,4) node[left] {\scriptsize{$-0.2$}};

    \fill (2.5,4.2) node[above] {\footnotesize{(c) $\mu_{0}=0.45$}};

    \draw[fill=red] (3.535,4) circle (2pt);
    
    \end{scope}
    
	\end{tikzpicture}
\end{center}\caption{This figure shows Proposer's indirect utility $\widetilde{U}(p)$ for different values of $\mu_{0}$ when $c$ is linear. The parameter values used for this figure are $\ell=0.15$ and $h=0.7$.
}
\label{fig:ExpectedPayoff}
\end{figure}

\cref{fig:ExpectedPayoff} illustrates \cref{prop:propose_first_optimal}. The optimal proposal is either $\underline{p}$ or $h$. It is certainly intuitive that $\psi(\mu_{0})$---the maximal proposal Vetoer would accept with no information---can be optimal. The potential optimality of $h$ is similar to the reason why Proposer never induces $\mu\in(0,\phi(h))$ in \cref{ex:persuade_first_linear}: When Vetoer's ideal point is $h$, an increase of $p$ is beneficial to him if $p<h$ but not if $p>h$. As reflected in the kinks of $\widetilde{U}(p)$ at $h$, this introduces a discontinuity in Proposer's incentive to raise $p$ at $h$, which makes $h$ a particularly tempting option. 

It is intriguing that $h$ is optimal---so vetoes occur with a positive probability---only when $\mu_{0}$ is neither too small nor too large. For $\mu_{0}$ sufficiently large, the result simply follows from the fact that $\underline{p}>h$, that is, Vetoer would be willing to accept even a higher proposal than $h$ with no information. To see why $\mu_{0}$ should not be too low, consider $p\in[2\ell,h)$. As explained above, an increase of $p$ is beneficial to a vetoer with ideal point $h$. However, it is still harmful if it is $\ell$. Therefore, Proposer's overall incentive to increase $p$ depends on the relative probability of ideal point $h$. If $\mu_{0}$ is sufficiently small then Proposer's incentive to raise $p$ is small and so $\underline{p}$ is optimal. Otherwise, Proposer has a strong incentive to raise $p$, so $h$ can be optimal. 

\subsection{Proposal First vs Persuasion First}

\begin{figure}
\begin{center}
\begin{tikzpicture}[scale=1]
		
	\draw[line width=0.5pt] (0,4.5)--(0,-0.5)--(6.5,-0.5)--(6.5,4.5); 
	
	\fill (6.5,-0.5) node[right] {\footnotesize{$\mu_{0}$}};
    \fill (6.5,-0.5) node[below] {\footnotesize{$1$}};
    \fill (0,-0.5) node[below] {\footnotesize{$0$}};
	\fill (0,4.5) node[above] {\footnotesize{$\overline{U}(\mu_{0})$}};

    \draw[dotted] (0,4.2)--(6.5,4.2); 
    \fill (0,4.2) node[left] {\footnotesize{$0$}};

    \fill (0,0) node[left] {\footnotesize{$1-2\ell$}};

    \draw[dotted] (1.3929,-0.5)--(1.3929,0.675);
    \draw[dotted] (2.3636,-0.5)--(2.3636,2.4);
    \draw[dotted] (4.1364,-0.5)--(4.1364,4.2);

    \draw[dotted] (0,0)--(6.5,0); 
    
    \fill (1.3929,-0.5) node[below] {\footnotesize{$\ell/h$}};
    \fill (2.3636,-0.5) node[below] {\footnotesize{$\phi(h)$}};
    \fill (4.1364,-0.5) node[below] {\footnotesize{$\phi(1)$}};

    \draw[line width=0.5pt,blue] plot[smooth]
    (0,0)--(0.046429,0.013043)--(0.092857,0.026471)--(0.13929,0.040299)--(0.18571,0.054545)--(0.23214,0.069231)--(0.27857,0.084375)--(0.325,0.1)--(0.37143,0.11613)--(0.41786,0.13279)--(0.46429,0.15)--(0.51071,0.1678)--(0.55714,0.18621)--(0.60357,0.20526)--(0.65,0.225)--(0.69643,0.24545)--(0.74286,0.26667)--(0.78929,0.28868)--(0.83571,0.31154)--(0.88214,0.33529)--(0.92857,0.36)--(0.975,0.38571)--(1.0214,0.4125)--(1.0679,0.44043)--(1.1143,0.46957)--(1.1607,0.5)--(1.2071,0.53182)--(1.2536,0.56512)--(1.3,0.6)--(1.3464,0.63659)--(1.3929,0.675);

    \draw[line width=0.5pt,blue] plot[smooth]
    (1.3929,0.675)--(1.4252,0.7325)--(1.4576,0.79)--(1.4899,0.8475)--(1.5223,0.905)--(1.5547,0.9625)--(1.587,1.02)--(1.6194,1.0775)--(1.6517,1.135)--(1.6841,1.1925)--(1.7165,1.25)--(1.7488,1.3075)--(1.7812,1.365)--(1.8135,1.4225)--(1.8459,1.48)--(1.8782,1.5375)--(1.9106,1.595)--(1.943,1.6525)--(1.9753,1.71)--(2.0077,1.7675)--(2.04,1.825)--(2.0724,1.8825)--(2.1048,1.94)--(2.1371,1.9975)--(2.1695,2.055)--(2.2018,2.1125)--(2.2342,2.17)--(2.2666,2.2275)--(2.2989,2.285)--(2.3313,2.3425)--(2.3636,2.4);

    \draw[line width=0.5pt,blue] plot[smooth]
    (2.3636,2.4)--(2.4227,2.46)--(2.4818,2.52)--(2.5409,2.58)--(2.6,2.64)--(2.6591,2.7)--(2.7182,2.76)--(2.7773,2.82)--(2.8364,2.88)--(2.8955,2.94)--(2.9545,3)--(3.0136,3.06)--(3.0727,3.12)--(3.1318,3.18)--(3.1909,3.24)--(3.25,3.3)--(3.3091,3.36)--(3.3682,3.42)--(3.4273,3.48)--(3.4864,3.54)--(3.5455,3.6)--(3.6045,3.66)--(3.6636,3.72)--(3.7227,3.78)--(3.7818,3.84)--(3.8409,3.9)--(3.9,3.96)--(3.9591,4.02)--(4.0182,4.08)--(4.0773,4.14)--(4.1364,4.2);

    \draw[line width=0.5pt,blue] (4.1364,4.2)--(6.5,4.2); 
    
    \draw[line width=0.5pt,brown,dashdotted] plot[smooth]
    (1.3929,0.675)--(1.4252,0.70293)--(1.4576,0.73187)--(1.4899,0.76187)--(1.5223,0.793)--(1.5547,0.82531)--(1.587,0.85888)--(1.6194,0.89379)--(1.6517,0.93011)--(1.6841,0.96792)--(1.7165,1.0073)--(1.7488,1.0485)--(1.7812,1.0914)--(1.8135,1.1362)--(1.8459,1.1832)--(1.8782,1.2323)--(1.9106,1.2838)--(1.943,1.3379)--(1.9753,1.3947)--(2.0077,1.4545)--(2.04,1.5174)--(2.0724,1.5839)--(2.1048,1.6541)--(2.1371,1.7283)--(2.1695,1.807)--(2.2018,1.8906)--(2.2342,1.9795)--(2.2666,2.0742)--(2.2989,2.1754)--(2.3313,2.2838)--(2.3636,2.4);

     \draw[line width=0.5pt,brown,dashdotted] plot[smooth]
    (0,0)--(0.046429,0.013043)--(0.092857,0.026471)--(0.13929,0.040299)--(0.18571,0.054545)--(0.23214,0.069231)--(0.27857,0.084375)--(0.325,0.1)--(0.37143,0.11613)--(0.41786,0.13279)--(0.46429,0.15)--(0.51071,0.1678)--(0.55714,0.18621)--(0.60357,0.20526)--(0.65,0.225)--(0.69643,0.24545)--(0.74286,0.26667)--(0.78929,0.28868)--(0.83571,0.31154)--(0.88214,0.33529)--(0.92857,0.36)--(0.975,0.38571)--(1.0214,0.4125)--(1.0679,0.44043)--(1.1143,0.46957)--(1.1607,0.5)--(1.2071,0.53182)--(1.2536,0.56512)--(1.3,0.6)--(1.3464,0.63659)--(1.3929,0.675);

    \draw[line width=0.5pt,brown,dashdotted] plot[smooth]
    (2.3636,2.4)--(2.4227,2.46)--(2.4818,2.52)--(2.5409,2.58)--(2.6,2.64)--(2.6591,2.7)--(2.7182,2.76)--(2.7773,2.82)--(2.8364,2.88)--(2.8955,2.94)--(2.9545,3)--(3.0136,3.06)--(3.0727,3.12)--(3.1318,3.18)--(3.1909,3.24)--(3.25,3.3)--(3.3091,3.36)--(3.3682,3.42)--(3.4273,3.48)--(3.4864,3.54)--(3.5455,3.6)--(3.6045,3.66)--(3.6636,3.72)--(3.7227,3.78)--(3.7818,3.84)--(3.8409,3.9)--(3.9,3.96)--(3.9591,4.02)--(4.0182,4.08)--(4.0773,4.14)--(4.1364,4.2);

    \draw[line width=0.5pt,brown,dashdotted] (4.1364,4.2)--(6.5,4.2); 

    \draw[line width=0.6pt,red,dashed] (0,0)--(4.1364,4.2)--(6.5,4.2); 

	\end{tikzpicture}
\end{center}\caption{\label{fig:power_info}This figure shows Proposer's expected payoffs as a function of the prior $\mu_{0}$ when $c$ is linear, $\ell\geq 0$, and $h>1/2$. Blue solid: the proposal-first model. Red dashed: the persuasion-first model. Brown dash-dotted: no information.
} 
\end{figure}
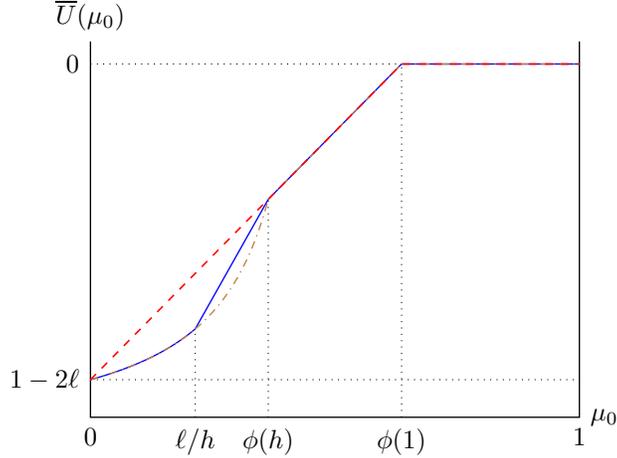

As is clear from \cref{fig:power_info}, in the current linear environment Proposer's expected payoff can be strictly larger under persuasion-first (red dashed) than under proposal-first (blue solid). To see why this difference arises, recall that in the proposal-first model, vetoes occur with a positive probability whenever Proposer chooses to provide information (with proposal $h$). In the persuasion-first model, Proposer can avoid vetoes even if she provides information and so induces a non-degenerate distribution of posteriors; if the realized posterior is $0$ then she can propose $2\ell$. This is not only directly beneficial to Proposer, but also enables her to be more aggressive with her information provision. 

\section{Conclusion}\label{sec:conclude} 

A central element of agenda-setting is the opportunity to persuade others to accept the proposal.  In this paper we provide the first analysis of the effect of persuasion in veto bargaining.  With quadratic loss we give a full characterization of Proposer's optimal strategy, showing that a simple binary experiment strictly increases the value of proposal power whenever the players are sufficiently misaligned.

We have analyzed one particular setting: when both players are initially symmetrically informed about Vetoer's optimal action.  An important avenue for future research is to understand how information provision affects bargaining in different informational environments. For example, in many settings the Proposer may be an expert and so may have information Vetoer lacks.  Alternatively, in a private values environment Vetoer's preferences are likely known only to him.  In parallel work we are exploring the effect of information provision in the latter setting.

\appendix

\section{Proofs}\label{appendix:proof}

\noindent 
\begin{proof}[Proof of \cref{prop:quadratic_persuade_first}]
    Consider a tangent line to $U$ at $(\mathbb{E}_{F}[\theta],U(\mathbb{E}_{F}[\theta]))$---the right-hand side of \eqref{eq:no_info_opt}. By Theorems 1 and 2 of \citet{DM:19}, the degenerate distribution $\delta_{\mathbb{E}_{F}[\theta]}$---so, no information---is optimal if and only if the tangent line stays uniformly above $U$. Due to the structure of $U$, this latter condition is equivalent to \eqref{eq:no_info_opt}. 

    If \eqref{eq:no_info_opt} fails then there exists a unique value of $s_{\ast}\in[\underline{\theta},0]$ that satisfies \eqref{eq:some_info_opt}; we establish this below. Given $s_{\ast}$ satisfying \eqref{eq:some_info_opt}, by Theorems 1 and 2 of \citet{DM:19}, a distribution $G$ is optimal if and only if $G(s_{\ast})=F(s_{\ast})$, $G$ puts all probability mass above $s_{\ast}$ on $\mathbb{E}[\theta|\theta\geq s_{\ast}]$, and the line that passes $(s_{\ast},U(s_{\ast}))$ and $(\mathbb{E}[\theta|\theta\geq s_{\ast}],U(\mathbb{E}[\theta|\theta\geq s_{\ast}])$ is uniformly larger than $U$ above $s_{\ast}$. The binary signal reveals whether $\theta\geq s_{\ast}$ or not satisfies all these properties and so is optimal to Proposer. 

Finally, we prove that there exists a unique value of $s_{\ast}\in[\underline{\theta},0]$ that satisfies \eqref{eq:some_info_opt}. For each $s_{\ast}<0$, let $s^{\ast}\in[0,1/2]$ be the value such that
    \begin{align*}
        U(s)\leq U(s_{\ast})+\frac{U(s^{\ast})-U(s_{\ast})}{s^{\ast}-s_{\ast}}(s-s_{\ast})\text{ for all }s\geq[s_{\ast},\overline{\theta}]. 
    \end{align*}
    The structure of $U$ ensures that $s^{\ast}$ is well defined for each $s_{\ast}\in[\underline{\theta},0)$. The desired value of $s_{\ast}$ is one such that $s^{\ast}=\mathbb{E}[\theta|\theta\geq s_{\ast}]$. The unique existence of such a value follows from the fact that (i) $s^{\ast}$ weakly decreases, while $\mathbb{E}[\theta|\theta\geq s_{\ast}]$ increases, in $s_{\ast}$ (so the two sides can be identical only at one value of $s_{\ast}$), (ii) $s^{\ast}>\mathbb{E}[\theta]=\mathbb{E}[\theta|\theta\geq s_{\ast}]$ if $s_{\ast}=\underline{\theta}$, and (iii) $s^{\ast}=s_{\ast}<\mathbb{E}[\theta|\theta\geq s_{\ast}]$ if $s_{\ast}$ is sufficiently close to $0$. 
\end{proof}

\noindent \begin{proof}[Proof of \cref{prop:proposal_first}]
    Recall that in \cref{prop:quadratic_persuade_first}, either no information is optimal, or any optimal signal reveals (effectively) only whether $\theta<s_{\ast}$ or not for some $s_{\ast}\in(\underline{\theta},0)$. No information (with proposal $p=2\mathbb{E}[\theta]$) can be obviously implemented in the proposal-first model. 
    
    Consider any signal that falls into the second category (revealing only whether $\theta<s_{\ast}$ or not). Under such a signal, the implemented policy is $2\mathbb{E}[\theta|\theta\geq s_{\ast}]$ if $\theta\geq s_{\ast}$ and the status quo policy $0$ if $s<s_{\ast}$. Suppose Proposer chooses the same experiment and offers $p=2\mathbb{E}[\theta|\theta\geq s_{\ast}]$. Clearly, Vetoer will accept $p$ if $\theta\geq s_{\ast}$ but vetoes if $\theta<s_{\ast}$, which is the same outcome as in the persuasion-first model. 
\end{proof}

\noindent \begin{proof}[Proof of \cref{p:statics}]
\underline{Part 1}:  Inequality \eqref{eq:some_info_opt} holds with equality when $s=s_*$, and so to hold for all $s<s_*$ and $s>s_*$ it must be that
\[
U'(s_*)=\frac{U(\mathbb{E}[\theta|\theta\geq s_{\ast}])-U(s_{\ast})}{\mathbb{E}[\theta|\theta\geq s_{\ast}]-s_{\ast}}, 
\]
or equivalently, 
\begin{equation}
\label{e:cutoff}
-s_*=\frac{u(2E[\theta|\theta \geq s_*])-u(0)}{2u'(2E[\theta|\theta \geq s_*])}-E[\theta|\theta \geq s_*].
\end{equation}
As the LHS is strictly decreasing in $s_*$ a change in $u$ which increases the RHS for all values of $s_*$ must decrease the solution to \eqref{e:cutoff}. So it is sufficient to show that for any $a=2E[\theta|\theta \geq s_*] \in [0, 1]$,
\begin{equation}
\label{e:inequality}
\frac{u_1(a)-u_1(0)}{u_1^{\prime}(a)} \leq \frac{u_2(a)-u_2(0)}{u_2^{\prime}(a)}.
\end{equation}
Now note that
\[
u_j(a)-u_j(0)=\int_{0}^{a} u_j^{\prime}(t) d(t),
\]
and so
\[
\frac{u_j(a)-u_j(0)}{u_j^{\prime}(a)}=\int_{0}^{a} \frac{u_j^{\prime}(t)}{u_j^{\prime}(a)} dt
\]
and
\[
\log \left(\frac{u_j^{\prime}(t)}{u_j^{\prime}(a)} \right) =\log (u_j^{\prime}(t))-\log(u_j^{\prime}(a))=\int_{t}^{a} -\frac{u_j^{\prime \prime}(s)}{u_j^{\prime}(s)} ds.
\]
Hence if
\[
-\frac{u_1^{\prime \prime}(a)}{u_1^{\prime}(a)} \leq -\frac{u_2^{\prime \prime}(a)}{u_2^{\prime}(a)}
\]
for all $a$ then \eqref{e:inequality} holds.  Thus $s_{*}$ and $s^*=E[\theta|\theta \geq s_*]$ are lower under $u_2$ than $u_1$.

\underline{Part 2}:  We first show that the RHS of \eqref{e:cutoff} is increasing in $s^*=E[\theta|\theta \geq s_*]$.  For this we note the the RHS is equal to
\[
\frac{u(2s^*)-u(0)-2u'(2s^*)s^*}{2u'(2s^*)}.
\]
Differentiating with respect to $s^*$, the derivative is proportional to
\begin{align*}
&2u'(2s^*)[2u'(2s*)-2u'(2s^*)-4s^*u^{\prime \prime}(2s^*)]-4u^{\prime \prime}(2s^*)[u(2s^*)-u(0)-2u'(2s^*)s^*] \\
&=-4u^{\prime \prime}(2s^*)[u(2s^*)-u(0)] \\
& \geq 0,
\end{align*}
and so the RHS is increasing.

A rightward LR shift increases $s^*=E[\theta|\theta \geq s_*]$ for any $s_*$. And so given the that LHS of \eqref{e:cutoff} is strictly decreasing in $s_*$, and the RHS is increasing, $s_*$ must decrease and $s^*=E[\theta|\theta \geq s_*]$ must increase for \eqref{e:cutoff} to continue to hold.  

\end{proof}

\noindent \begin{proof}[Proof of \cref{prop:no_info_optimal}]
We first prove that no information is optimal if $\mu_{0}\geq\max\{0,\phi(h)\}$. If $\mu_{0}\geq \phi(1)$ (equivalently, $\psi(\mu_{0})=1$) then the result is trivial because Vetoer will accept 1 with no information. In what follows, suppose $\mu_{0}<\phi(1)$. Let $\overline{U}$ denote the \emph{smallest} concave function uniformly above $U$. Since $\psi(\mu)\leq 2((1-\mu)\ell+\mu h)$ for all $\mu$ and $-c(1-2((1-\mu)\ell+\mu h))$ is concave in $\mu$, we have
\begin{equation*}
   -c(1-2((1-\mu)\ell+\mu h))\geq \overline{U}(\mu)\geq U(\mu)=-c(1-\psi(\mu)).
\end{equation*}
For $\mu\geq\max\{0,\phi(h)\}$, $\psi(\mu)=2((1-\mu)\ell+\mu h)$, and thus $\overline{U}(\mu)=U(\mu)$. Since $\overline{U}$ represents the maximal attainable payoff, it follows that no information is optimal whenever $\mu_{0}\geq\max\{0,\phi(h)\}$. 

For the second result, notice that the straight line that connects between $(0,U(0))$ and $(\phi(h),U(\phi(h)))$ necessarily lies weakly below $\overline{U}(\mu)$. Therefore, a \emph{sufficient} condition for no information to be suboptimal at $\mu$ close to $0$ is 
\begin{equation*}
    U^{\prime}(0)<\frac{U(\phi(h))-U(0)}{\phi(h)}\leq\overline{U}^{\prime}(0).
\end{equation*}    
Applying the above derivation, we can show that this inequality is equivalent to
\begin{equation*}
    c^{\prime}(1-\psi(0))\psi^{\prime}(0)=c^{\prime}(1-2\ell)2\ell<(c(1)-c(1-h))\frac{2(h-\ell)}{h-2\ell},
\end{equation*}
which necessarily holds when $\ell$ is close to $0$. 
\end{proof}

\section{Example: More Than Two Messages}  
\label{three}


We consider the persuasion-first model in which both Proposer and Vetoer have linear loss functions; $\theta$ can be either $0$, $\ell=0.1$, or $h=0.5$; and the prior probabilities of Vetoer's ideal point are given by $(0.7,0.2,0.1)$. In this case, Proposer obtains the following expected payoff by providing full information (and optimally proposing $0$, $0.2$, or $1$, depending on $\theta$):
\[
U_{FI}=0.7(-1)+0.2(-0.8)+0.1(0)=-0.86.
\]
We now show that no binary experiment can achieve the same payoff.

\paragraph{Optimal proposal depending on Vetoer's posterior belief.} If Vetoer's belief over $\theta$ is given by $(\mu_0, \mu_{\ell}, 1-\mu_0-\mu_{\ell})$ then he is willing to accept $p$ if and only if 
\begin{align*}
-\mu_0 p-\mu_{\ell}|p-0.1|-(1-\mu_0-\mu_{\ell})|p-0.5|\geq -\mu_{\ell}0.1-(1-\mu_0-\mu_{\ell})0.5.   \end{align*}
The highest $p$ that Vetoer would accept (equivalently, Proposer's optimal proposal) is given by
\begin{align}
\label{action}
p(\mu_0, \mu_l):=
\left\{\begin{array}{ll}
    0 & \text{if } \mu_0 > 1/2 \\
    \frac{0.2\mu_{\ell}}{2(\mu_0+\mu_{\ell})-1} & \text{if } \mu_0 \in (1/2-0.8 \mu_{\ell}, 1/2] \\
    1-\mu_0-0.8\mu_{\ell} & \text{if } 2\mu_0+1.6 \mu_{\ell} \leq 1. 
\end{array}\right.
\end{align}
Notice that if Proposer provides no information then, since the prior probability of $\theta=0$ is $0.7$, Vetoer would never accept any positive proposal, so Proposer's expected payoff is 
\[
U_{NO}=-1<U_{FI}.
\]

\paragraph{Optimal binary experiment.} Consider the following binary experiment: If $\theta=0$ then $s=0$ with probability $1$. If $\theta=0.1$ then $s=0$ with probability $1/6$ and $s=1$ with probability $5/6$. Finally, if $\theta=0.5$ then $s=1$ with probability $1$. Together with the prior $(0.7,0.2,0.1)$, this leads to posterior $(21/22,1/22,0)$ with probability $11/15$ (when $s=0$) and $(0,5/8,3/8)$ with probability $4/15$ (when $s=1$). The derivation above implies that Proposer will propose $0$ in the former case and $1/2$ in the latter case. The resulting expected payoff is
\begin{align*}
    U_{BI}=\frac{11}{15}(-1)+\frac{4}{15}(-0.5)=-\frac{13}{15} \approx -0.8667<U_{FI}=-0.86.
\end{align*}

To understand this result, notice that analogous to the binary analysis (\cref{ex:persuade_first_linear}), Proposer optimally induces beliefs for which proposal $p=h=0.5$ is made and accepted; this is the reason why this experiment is an optimal binary experiment, as shown shortly. With only two messages, however, the remaining $\ell=0.1$ vetoers must be pooled with $0$, at the cost of maintaining the status quo more often.

We conclude this section by proving that the above binary experiment is an optimal experiment. To begin with, notice that at least one message (signal realization) should have $\mu_{0}\geq 0.7>0.5$, in which case the induced policy is $0$. Hence, without loss, we can assume that $s\in\{0,p\}$ where $s$ denotes the policy induced after each realization. 


Let $\sigma(\theta) \in [0, 1]$ denote the probability of $s=p$ given $\theta$. Since $p$ is at least weakly higher with more weight on $0.5$ than $0.1$, and at least weakly higher with more weight on $0.1$ than $0$, it incurs no loss of generality to restrict attention to $\sigma(0.5)=1$ and either $\sigma(0)=0$ or $\sigma(0.1)=1$. In other words, $p$ (as opposed to $0$) should be induced with probability $1$ if $\theta=0.5$. In addition, $p$ should be induced first when $\theta=0.1$ and then (possibly) when $\theta=0$; if $\sigma(0.1)<1$ while $\sigma(0)>0$ then, for example, $p$ can be raised without reducing the probability of inducing it by raising $\sigma(0.1)$ while lowering $\sigma(0)$. 


\textbf{(i) When $\sigma(0.1)=1$:} Consider the case in which $\sigma(0.1)=1$ and $\sigma(0)=\sigma \in [0, 1]$.  Then $s=p$ is chosen with probability $0.3+0.7\sigma$ and $0$ is chosen with probability $0.7(1-\sigma(0))$. The posterior following $s=p$ is then 
\[
\mu_0=\frac{7\sigma}{3+7\sigma}\text{ and }\mu_{\ell}=\frac{2}{3+7\sigma}.
\]
Obviously any $\sigma$ for which $\mu_0 >1/2$ cannot be optimal, and there is no $\sigma$ for which  
$2\mu_0+1.6 \mu_{\ell}=\frac{3.2 +14\sigma}{3+7\sigma} \leq 1$. This implies that the resulting action is determined by the second case in \eqref{action}:
\[
p=\frac{0.2\mu_{\ell}}{2(\mu_0+\mu_{\ell})-1}=\frac{0.4}{1+7\sigma}.
\]
Then, Proposer's payoff for any $\sigma (0) \in [0, 3/7]$ is 
\[
-0.7(1-\sigma)-(0.3+0.7\sigma)\left(1-\frac{0.4}{1+7\sigma}\right)=-1+\frac{0.4(0.3+0.7\sigma)}{1+7\sigma},
\]
which is decreasing in $\sigma(0)$. Hence the optimal experiment in this class is $\sigma(0)=0$, in which case Proposer's expected payoff is given by
\begin{equation*}
U_{BI}^{\prime}=-1+(0.4)(0.3)=-0.88
\end{equation*}

\textbf{(ii) When $\sigma(0)=0$:} Now consider the other case in which $\sigma(0)=0$ and $\sigma(0.1)=\sigma \in [0, 1]$.  Message $p$ then occurs with probability $0.1+0.2\sigma$. As $\mu_0=0$ conditional on $s=p$, it follows from \eqref{action} that 
\begin{align*}
p(0,\mu_{\ell})=
\left\{\begin{array}{cc}
    \frac{0.2\mu_{\ell}}{2\mu_{\ell}-1} & \text{if } \mu_{\ell} > 2/5 \\
    1-0.8\mu_{\ell} & \text{if } \mu_{\ell} \leq 2/5. 
\end{array}\right.
\end{align*}
Combining this with $\mu_{\ell}=\frac{2\sigma}{2\sigma+1}$ leads to
\begin{align*}
p=
\left\{\begin{array}{cc}
    \frac{0.4\sigma}{2\sigma-1} & \text{if } \sigma > 5/6 \\
    \frac{1.4\sigma+1}{2\sigma+1} & \text{if } \sigma \leq 5/6. 
\end{array}\right.
\end{align*}
Consequently, Proposer's expected utility is 
\begin{align*}
-(0.9-0.2\sigma)(1)-(0.1+0.2\sigma)(1-p)&=-1+(0.1+0.2\sigma)p\\
&=\left\{\begin{array}{cc}
    -1+(0.1+0.2\sigma)\frac{0.4\sigma}{2\sigma-1} & \text{if } \sigma > 5/6 \\
    -1+(0.1+0.2 \sigma)\frac{0.4\sigma+1}{2\sigma+1} & \text{if } \sigma \leq  5/6. 
\end{array}\right.
\end{align*}
This is maximized when $\sigma=5/6$ at which point $p=0.5$ and Proposer's utility is $U_{BI}=-13/15$. The optimality of this binary experiment follows from the fact that $U_{BI}>U_{BI}^{\prime}$.

\clearpage
\bibliography{veto_info_references}

\end{document}